\documentclass[12pt]{iopart}
\usepackage{epsfig}
\usepackage{graphicx}

\newcommand{\here}{\makebox(0,0)}
\newcommand{\room}{\rule[-0.3cm]{0cm}{0.8cm}}
\newcommand{\be}{\begin{equation}}
\newcommand{\ee}{\end{equation}}
\newcommand{\bd}{\begin{displaymath}}
\newcommand{\ed}{\end{displaymath}}
\newcommand{\vsp}{\vspace*{3mm}}
\newcommand{\pprime}{{\prime\prime}}

\newcommand{\Bra    }{\left\langle}
\newcommand{\Ket    }{\right\rangle}
\newcommand{\bra    }{\langle}
\newcommand{\ket    }{\rangle}

\newcommand{\order}{{\cal O}}

\newcommand{\bnull}{\mbox{\boldmath $0$}}
\newcommand{\R}{{\rm I\!R}}
\newcommand{\bsigma}{{\mbox{\boldmath $\sigma$}}}

\newcommand{\bmu}{{\mbox{\boldmath $\mu$}}}

\newcommand{\he}{\hat{\mathbf{e}}}
\newcommand{\bx}{\mbox{\boldmath $x$}}

\newcommand{\param}{\bmu}

\newcommand{\by}{\ensuremath{\mathbf{y}}}

\newcommand{\bU}{\ensuremath{\mathbf{U}}}

\newcommand{\btau}{{\mbox{\boldmath $\tau$}}}

\newcommand{\unitv}{{\mbox{\boldmath $e$}}}

 \newcommand{\one}{{\rm 1\!\!I}}
 \newcommand{\N}{{\rm I\!N}}

\begin{document}

\title[Finitely connected vector spin systems with random matrix interactions]{Finitely connected vector spin systems with random matrix interactions}

\author{ACC Coolen$^\dag$, NS Skantzos$^\S$, I P\'{e}rez Castillo$^\ddag$,  CJ P\'erez Vicente$^\star$,
JPL Hatchett$^\diamondsuit$, B Wemmenhove$^\P$ and T
Nikoletopoulos$^\dag$}
\address{\dag ~ Department of Mathematics, King's College London, The Strand,
London WC2R 2LS, United Kingdom}
\address{\S~ Institute for Theoretical Physics, Celestijnenlaan 200D, Katholieke Universiteit Leuven, B-3001 Belgium}
\address{\ddag~ Rudolf Peierls Center for Theoretical Physics, University of Oxford, 1 Keble Road, Oxford, OX1 3NP, United Kingdom}
\address{$\star$~ Departament de F\'isica Fonamental, Facultat de F\'isica,
 Universitat de Barcelona, 08028 Barcelona, Spain}
 \address{$\diamondsuit$~Laboratory for Mathematical Neuroscience, RIKEN
 Brain Science Institute, Hirosawa 2-1, Wako-Shi, Saitama
 351-0198, Japan}
\address{\P~ Department of Medical Physics and Biophysics,
Radboud University Nijmegen, Geert Grooteplein 21, NL 6525 EZ Nijmegen, The Netherlands}

\begin{abstract}
We use finite connectivity equilibrium replica theory to solve
models of finitely connected unit-length vectorial spins, with
random pair-interactions which are of the orthogonal matrix type.
Since the spins are continuous and the connectivity $c$ remains
finite in the thermodynamic limit, the replica-symmetric order
parameter is a functional. The general theory is developed for
arbitrary values of the dimension $d$ of the spins, and arbitrary
choices of the ensemble of random orthogonal matrices. We
calculate phase diagrams and the values of moments of the order
parameter explicitly for  $d=2$ (finitely connected XY spins with
random chiral interactions) and for $d=3$ (finitely connected
classical Heisenberg spins with random chiral interactions).
Numerical simulations are shown to support our predictions quite
satisfactorily.
\end{abstract}

 \ead{\tt
tcoolen@mth.kcl.ac.uk,  Nikos.Skantzos@fys.kuleuven.ac.be,
isaac@thphys.ox.ac.uk, conrad@ffn.ub.es, hatchett@brain.riken.jp,
B.Wemmenhove@science.ru.nl, theodore@mth.kcl.ac.uk}

\section{Introduction}

Models of finitely connected disordered spin systems have been
studied for some twenty years, following the initiating papers
\cite{viana-bray85,kanter-sompo87,mezard-parisi87,mottishaw-dedominicis87,wong-sherrington88}.
Especially due to the unexpectedly rich and varied range of
multi-disciplinary applications of finite connectivity replica
techniques which emerged subsequently, in e.g. spin-glass modeling
\cite{monasson98,mezard-parisi01,mezard-parisi03,mini-guzai},
 error correcting codes
\cite{murayama00,nakamura,nishimori,nikos}, theoretical computer
science
\cite{kirkpatrick-selman94,monasson-zecchina98,monasson-zecchina982,monasson-zecchina99},
recurrent neural networks
\cite{wemmenhove-coolen03,perez-skantzos03,guzai3}, and
`small-world' networks \cite{guzai2}, this field is presently
enjoying a renewed interest and popularity. Until very recently,
analysis was limited to the equilibrium properties of such models,
but now attention has also turned to  the dynamics of finitely
connected spin systems \cite{semerjian1,semerjian2,guzai1,DRT},
using combinatorial and  generating functional methods. In the
domain of physical spin systems, research into finitely connected
systems has usually been triggered by the desire to develop
solvable spin-glass models which are closer to real
finite-dimensional systems than the celebrated fully connected
spin-glass model of \cite{SK}. As far as we are aware, however
(and in contrast to the situation with fully connected disordered
spin systems), all finitely connected and disordered spin systems
analyzed theoretically so far involved scalar spin variables
(either of the Ising type, the soft-spin type, or the spherical
type).

In the present paper we solve equilibrium models of finitely
connected spin systems of unit-length vectorial spins, and with
random pair-interactions between them  which are of a chiral
nature, defined by random orthogonal matrices which promote random
relative spatial orientations between pairs of spins. The
motivation behind our study is twofold. Firstly, we aim to expand
the domain of solvable and solved finitely connected spin models,
by also including those where the spins have a truly vectorial
character, and where their interactions are of a (random) matrix
type. Vectorial spins are not only more realistic from a
fundamental physical point of view in any magnetic system, but are
also of special relevance in the context of e.g. Josephson
junctions \cite{Wies,PB222a,PB222b}. In these latter junctions it
is essential for the spin-interactions to have a chiral character.
The inclusion of bond- and field-disorder in chiral spin systems
has so far only been studied analytically in a mean-field setting,
where all spins are allowed to interact with each other (purely
for mathematical convenience,  see e.g. \cite{CP} and references
therein). The present study can be regarded as a step away from
the unrealistic full connectivity towards finite dimensional
models in such vectorial systems. Our second motivation is of a
technical nature. In contrast to models with discrete (e.g. Ising)
spins, when choosing real-valued microscopic variables the
replica-symmetric (RS) theory will involve an order parameter
which is itself a {\em functional}, rather than a function.
Solving the associated order parameter equations is therefore
non-trivial, and leads to many numerical complications especially
when the domain of values for the individual spins is not bounded
(as is the case for e.g. soft spins). In our present model we have
vectorial spin states; although  continuous and therefore leading
to a theory involving an order parameter functional, each spin
state represents a point on a sphere and has therefore a compact
domain. This is found to be a considerable mathematical and
numerical advantage, and allows us to push our analysis and
therefore also our understanding  further.

The structure of our paper is as follows. We first develop the
general replica symmetric theory for finitely connected vectorial
spin systems with random chiral interactions described by
orthogonal matrices, for arbitrary dimensions of the microscopic
sphere $S_{d-1}$ which constrains the values of the individual
spins. We then apply our theory first to the case $d=2$, where the
spins reduce to XY ones, followed by application to $d=3$ where we
have classical Heisenberg spins. In both cases we calculate phase
diagrams (with continuous transition lines calculated via
functional moment expansions) as well as the values of macroscopic
observables (using truncated population dynamics routines), for
different choices for the ensemble of random orthogonal matrices.
We complement our results with numerical simulations. The latter
are found to agree well with our theoretical predictions,
especially if the usual problems associated with the
moment-truncation of population dynamics algorithms and the
well-known finite size effects in disordered spin systems are
taken into account.

\section{Definitions}

We study finitely connected and bond-disordered systems composed
of $N$ interacting normalized vectorial soft spins $\bsigma_i\in
S_{d-1}$, with $S_{d-1}$ denoting the unit sphere in $\R^{d}$.
Thus for $d=2$ our spins become XY spins, with each spin
representing a point on a the unit circle, whereas for $d=3$ they
become classical Heisenberg spins, with each representing a point
on the unit sphere, etc. Our systems are taken to be in thermal
equilibrium, characterized by Hamiltonians  of the following form,
with the short-hand $\{\bsigma\}=(\bsigma_1,\ldots,\bsigma_N)$:
\be
H(\{\bsigma\})=-J\sum_{i<j}c_{ij}\bsigma_i\cdot
\bU_{ij}\bsigma_j+\sum_i V(\bsigma_i) \label{eq:Hamiltonian} \ee
  The independently
distributed quenched random variables $c_{ij}\in\{0,1\}$ define
the (random)  connectivity of the system, i.e. they dictate which
pairs of spins are allowed to interact. The real $d\times d$
matrices are assumed orthogonal and with determinant one, i.e. they
represent rotations in $\R^{d}$, and are drawn randomly and
independently from some random matrix ensemble characterized by a
distribution $P(\bU)$. We will assume as yet only that
$P(\bU)=P(\bU^\dag)$. The single-site potentials $V(\bsigma)$
serve
to break symmetries, and will also enable us to take an
Ising limit later as a benchmark test (e.g. for
$V(\bsigma)=\gamma(\hat{\unitv}\cdot\bsigma)^2$, with
$\hat{\unitv}$ denoting a fixed unit length vector in $\R^d$, and
where $\gamma\to\infty$). In this paper we will be concerned with
the so-called finite connectivity regime, where
\be
P(c_{ij})=\frac{c}{N}\delta_{c_{ij},1}+(1-\frac{c}{N})\delta_{c_{ij},0}~~~~{\rm
for~all}~~i<j
\label{eq:connectivity}
 \ee
 with $c=\order(N^0)$. In this limit, each spin interacts on average only with a {\em finite} number $c$ of other spins, even in the thermodynamic limit
 (similar to finite dimensional systems).
 In this sense, in spite of the absence of geometrical
 considerations, finitely connected spin models can be regarded as
 closer to physical reality than fully connected ones.

 In the remainder of this paper,  averages over the random connectivity $\{c_{ij}\}$ and over the random
ortoghonal matrices $\{\bU_{ij}\}$ will be denoted by
$\overline{\cdots}$.
 We will use the standard replica
identity $\overline{\log Z}=\lim_{n\to
0}n^{-1}\log\overline{Z^n}$,  with $\overline{Z^n}$ initially
evaluated for integer $n$, to calculate for our system the
asymptotic disorder-averaged free energy per spin
$\overline{f}=-\lim_{N\to\infty}(\beta N)^{-1}\overline{\log Z}$.
Indices will be used according to the standard conventions, with
Greek ones labeling replicas ($\alpha=1,\ldots,n$) and Roman ones
labeling spins ($i=1,\ldots,N$).

\section{Replica calculation of the disorder-averaged free energy
per spin}

\subsection{Derivation of replica saddle-point equations}

In evaluating the disorder-averaged free energy per spin with the
replica method,
\begin{equation}
\overline{f} = -\lim_{N\to\infty}\frac{1}{\beta N}
\overline{\left[\log \int_{S_{d-1}}\![\prod_i
\rmd\bsigma_i]~e^{-\beta H(\{\bsigma\})}\right]}
\label{eq:define_f}
\end{equation}
 one finds that site factorization can be
achieved upon isolating the usual site-averaged replica order
parameter for finitely connected systems. After  redefining
$\{\bsigma\}=(\bsigma^1,\ldots,\bsigma^n)$, where now
$\bsigma_i^\alpha,\bsigma^\alpha \in S_{d-1}$, this order
parameter takes the form
$P(\{\bsigma\})=\lim_{N\to\infty}N^{-1}\sum_i
\overline{\prod_\alpha\delta[\bsigma^\alpha-\bsigma^\alpha_i]}$.
The insertion of appropriate functional $\delta$-distributions
into (\ref{eq:define_f}) gives
 \begin{eqnarray*}
 \hspace*{-10mm}
\overline{f}&=& -\lim_{N\to\infty}\lim_{n\to 0}\frac{1}{\beta
Nn}\log\int\!\!\ldots\!\!\int_{S_{d-1}}\!\Big[\prod_{i\alpha}\rmd\bsigma^\alpha_i
e^{-\beta V(\bsigma^\alpha_i)}\Big] \prod_{i<j}\overline{e^{\beta
J c_{ij}\sum_\alpha \bsigma^\alpha_i\cdot
\bU_{ij}\bsigma^\alpha_j}} \nonumber\\
 \hspace*{-10mm} &=&
-\lim_{N\to\infty}\lim_{n\to 0}\frac{1}{\beta
Nn}\log\int\!\!\ldots\!\!\int_{S_{d-1}}\Big[\prod_{i\alpha}
\rmd\bsigma^\alpha_i e^{-\beta V(\bsigma^\alpha_i)}\Big]
 \nonumber
\\
 \hspace*{-10mm}
 & &  \times \exp\left\{ \frac{c}{2N}\sum_{ij}\Big[
\int\!\rmd\bU~P(\bU)  e^{\beta J\sum_\alpha \bsigma^\alpha_i\cdot
\bU \bsigma^\alpha_j}-1\Big]+\order(N^{0})
 \right\}
 \nonumber\\
 \hspace*{-10mm}
 &=&-\lim_{N\to\infty}\lim_{n\to 0}\frac{1}{\beta
Nn}\log\int\!\left\{ \prod_{\{\bsigma\}}\frac{\rmd\!
P(\{\bsigma\})\rmd\hat{P}(\{\bsigma\})e^{\rmi N
P(\{\bsigma\})\hat{P}(\{\bsigma\})}}{2\pi/N}\right\} \nonumber
\\
\hspace*{-10mm}
 & &\times \exp\left\{ \frac{1}{2}cN\int\!\{\rmd\bsigma
\rmd\bsigma^\prime\} ~P(\{\bsigma\})P(\{\bsigma^\prime\})[
\int\!\rmd\bU~P(\bU) e^{\beta J \sum_\alpha\bsigma_\alpha\cdot\bU
\bsigma_\alpha^\prime}-1]
 \right\}
 \nonumber
\\
\hspace*{-10mm} && \times \exp\left\{N\log \int\!\{\rmd\bsigma\}~
e^{-\rmi\sum_{\{\bsigma^\prime\}}
\hat{P}(\{\bsigma^\prime\})\prod_\alpha\delta[\bsigma^{\alpha
\prime} -\bsigma^\alpha]-\beta \sum_\alpha
V(\bsigma^\alpha)}\right\}
 \nonumber
\end{eqnarray*}
In this expression  we substitute $\hat{P}(\{\bsigma\})\to
\{\rmd\bsigma\} \hat{P}(\{\bsigma\})$; this subsequently enables
us to
 take a continuum limit, whereby $\{\rmd\bsigma\}\to \{\bnull\}$.
Upon writing the resulting path integration measure as
$\prod_{\{\bsigma\}}[\rmd\!
P(\{\bsigma\})\rmd\hat{P}(\{\bsigma\})/2\pi]=\{\rmd\! P
\rmd\hat{P}\}$, and upon neglecting irrelevant constants, our
expression for the disorder averaged free energy per spin is then
seen to take a saddle-point form:
 \begin{eqnarray}
 \hspace*{-15mm}
\overline{f} &=&
  -\lim_{N\to\infty}\lim_{n\to 0}\frac{1}{\beta
Nn}\log\int\!\{\rmd\! P\rmd\hat{P}\}~ e^{\rmi N
\int\!\{\rmd\bsigma\}~ P(\{\bsigma\})\hat{P}(\{\bsigma\})}
\nonumber
\\
 \hspace*{-15mm}
&& \times \exp\left\{ \frac{1}{2}cN\int\!\{\rmd\bsigma
\rmd\bsigma^\prime \}~P(\{\bsigma\})P(\{\bsigma^\prime\})\Big[
\int\!\rmd\bU~P(\bU) e^{\beta J \sum_\alpha \bsigma^\alpha\cdot\bU
\bsigma^{\alpha\prime}}\!\!-1\Big]
 \right\}
 \nonumber
\\
 \hspace*{-15mm}
&& \times \exp\left\{ N\log \int\!\{\rmd\bsigma\}~ e^{-\beta
\sum_{\alpha} V(\bsigma_\alpha)-\rmi \hat{P}(\{\bsigma\})}
\right\}
 \nonumber
 \\
  \hspace*{-15mm}
 &=&
  -\lim_{n\to 0}\frac{1}{\beta
n}{\rm extr}_{\{P,\hat{P}\}}\left\{ \rmi\!\int\!\{\rmd\bsigma\}~
P(\{\bsigma\})\hat{P}(\{\bsigma\}) +\log \int\!\{\rmd\bsigma\}~
e^{-\beta \sum_{\alpha} V(\bsigma_\alpha)-\rmi
\hat{P}(\{\bsigma\})} \right.
 \nonumber
\\
 \hspace*{-15mm}
&& \left. +\frac{1}{2}c\int\!\{\rmd\bsigma \rmd\bsigma^\prime\}
~P(\{\bsigma\})P(\{\bsigma^\prime\})\Big[ \int\!\rmd\bU~P(\bU)
e^{\beta J \sum_\alpha\bsigma^\alpha\cdot\bU
\bsigma^{\alpha\prime}}\!\!-1\Big]
 \right\}
 \label{eq:f_1}
\end{eqnarray}
The functional variation of (\ref{eq:f_1}) with respect to the two
order parameter functions $\hat{P}(\{\bsigma\})$ and
$P(\{\bsigma\})$ gives us the following two saddle-point
equations, respectively:
\be
P(\{\bsigma\}) =\frac{ e^{-\beta \sum_{\alpha}
V(\bsigma_\alpha)-\rmi \hat{P}(\{\bsigma\})}}
{\int\{d\bsigma^\prime\}~ e^{-\beta \sum_{\alpha}
V(\bsigma^{\alpha \prime})-\rmi \hat{P}(\{\bsigma^\prime\})}}
 \label{eq:SP1}
 \ee
\be
\hat{P}(\{\bsigma\}) = \rmi c\int\!\{\rmd\bsigma^\prime\}
~P(\{\bsigma^\prime\})\Big[ \int\!\rmd\bU~P(\bU) e^{\beta J
\sum_\alpha\bsigma^\alpha\cdot\bU
\bsigma^{\alpha\prime}}\!\!-1\Big] \label{eq:SP2}
 \ee
 Elimination of the conjugate order parameter function $\hat{P}(\{\bsigma\})$
 leads to a saddle-point equation for $P(\{\bsigma\})$ only, and
an associated  expression for the free energy $\overline{f}$:
\be
\hspace*{-10mm} P(\{\bsigma\}) = \frac{
e^{c\int\!\{\rmd\bsigma^\prime\} ~P(\{\bsigma^\prime\})[
\int\!\rmd\bU~P(\bU) e^{\beta J \sum_\alpha\bsigma^\alpha\cdot
\bU\bsigma^{\alpha \prime}}\!\!-1]-\beta \sum_{\alpha}
V(\bsigma^\alpha)}} {\int\!\{\rmd\bsigma^\prime\}~
e^{c\int\!\{\rmd\bsigma^\pprime\} ~P(\{\bsigma^\pprime\})[
\int\!\rmd\bU~P(\bU) e^{\beta J
\sum_\alpha\bsigma^{\alpha\prime}\cdot\bU
\bsigma^{\alpha\pprime}}\!\!-1]-\beta \sum_{\alpha}
V(\bsigma^{\alpha\prime})}}
 \label{eq:RSB_SPE}
 \ee
 \begin{eqnarray}
\hspace*{-10mm}
  \overline{f}
 &=&
  \lim_{n\to 0}\frac{1}{\beta n}
\left\{ \frac{1}{2}c\int\!\{\rmd\bsigma d\bsigma^\prime\}
~P(\{\bsigma\})P(\{\bsigma^\prime\})\Big[ \int\!\rmd\bU~P(\bU)
e^{\beta J \sum_\alpha\bsigma^\alpha\cdot\bU
\bsigma^{\alpha\prime}}\!\!-1\Big] \right. \nonumber
\\
\hspace*{-10mm} && \left.\hspace*{0mm}
 -\log \int\!\{\rmd\bsigma\}~ e^{c\int\!\{\rmd\bsigma^\prime\}
~P(\{\bsigma^\prime\})[ \int\!\rmd\bU~P(\bU) e^{\beta J
\sum_\alpha\bsigma^\alpha\cdot\bU
\bsigma^{\alpha\prime}}\!\!-1]-\beta \sum_{\alpha}
V(\bsigma^\alpha)}
 \right\}
 \label{eq:RSB_f}
\end{eqnarray}

\subsection{Replica symmetric theory}

We now make the canonical RS ansatz for continuous spins in our
saddle-point equations\footnote{This ansatz reflects the
complicating fact that, in the case of continuous spins, the RS
order parameter function depends on the replicated spin variables
not only  via the sum $\sum_\alpha \sigma_\alpha$ (as would have
been the case for Ising spins), but rather on all possible sums of
the form $\sum_{\alpha}\sigma_{\alpha}^K$, for any $K\geq 1$.}. We
assume there to be a complete family of distributions
$P[\bsigma|\param]$ on $S_{d-1}$, parametrized by a countable set
of real-valued parameters $\param=(\mu_0,\mu_1,\mu_2,\ldots)$,
such that
 \be
P_{\rm RS}(\bsigma^1,\ldots,\bsigma^n) =\int\!
\rmd\param~w(\param) \prod_{\alpha}P[\bsigma^\alpha|\param]
\label{eq:P_RS}
 \ee
  with a normalized density
 $w(\param)$.
A representation-independent but mathematically equivalent
formulation of the RS ansatz follows upon defining the functional
measure
\be
W[\{P\}]=\int\! \rmd\param~w(\param) \prod_{\bsigma\in S_{d-1}}
\delta\left[P(\bsigma)-P[\bsigma|\param]\right]
 \label{eq:functional_measure}
\ee
 For non-degenerate parametrizations, i.e. for those such that every function $P(\bsigma)$ corresponds to a
 unique choice
$\param(\{P\})$ of parameters, with
$P[\bsigma|\param(\{P\})]=P(\bsigma)$ for all $\bsigma\in
S_{d-1}$, we can invert relation (\ref{eq:functional_measure}) and
write
\be
w(\param)=\int\!\{\rmd\! P\}~W[\{P\}]~\delta[\param-\param(\{P\})]
\label{eq:inversion} \ee
 In terms of the functional measure $W[\{P\}]$
our RS ansatz (\ref{eq:P_RS}) takes an elegant and
representation-free form:
 \be
P_{\rm RS}(\bsigma^1,\ldots,\bsigma^n) =\int\!\{\rmd\!
P\}~W[\{P\}]~ \prod_{\alpha}P(\bsigma_\alpha)
\label{eq:RS_alternative}
 \ee
The physical interpretation of our subsequent observables and
results in terms of the original disordered $N$-spin system will
follow from the identity
\be
\int\!\{\rmd\! P\}~W[\{P\}]\prod_\alpha
\left[\int\!\rmd\bsigma~P(\bsigma)f_\alpha(\bsigma)\right]=\lim_{N\to\infty}\frac{1}{N}\sum_i\overline{\prod_\alpha
\bra f_\alpha(\sigma_i)\ket} \label{eq:physical_meaning}
 \ee
We insert into our general saddle-point equation
(\ref{eq:RSB_SPE}) the RS ansatz (\ref{eq:RS_alternative}), and
introduce the convention $\prod_{k=1}^0 a_k=1$ for any series
$\{a_k\}$. This leads to the following identity (with a  constant
$C_n$ which will in due course be determined by normalization):
\begin{eqnarray}
\hspace*{-25mm} C_n \int\!\{\rmd\! P\}~W[\{P\}]
\prod_{\alpha}P(\bsigma_\alpha) &&\nonumber \\ \hspace*{-20mm} &&
 \hspace*{-20mm} = e^{c\int\!\{\rmd\! P\}
W[\{P\}]\int\! \rmd\bU~ P(\bU)\prod_\alpha \left[
\int\!\rmd\bsigma^\prime P(\bsigma^\prime) e^{\beta J
\bsigma_\alpha\cdot \bU \bsigma^\prime}\right]-c-\beta \sum_\alpha
V(\bsigma_\alpha)}\nonumber
\\
\hspace*{-20mm} && \hspace*{-20mm} = \sum_{\ell\geq
0}\frac{c^\ell}{\ell !}e^{-c} \int\!\prod_{k=1}^\ell[\{\rmd\!
P_k\} W[\{P_k\}]\rmd\bU_k P(\bU_k)] \nonumber
\\
\hspace*{-20mm} && \hspace*{-15mm}\times \prod_\alpha
\left\{e^{-\beta V(\bsigma_\alpha)} \prod_{k=1}^\ell
\int\!\rmd\bsigma^\prime P_k(\bsigma^\prime) e^{\beta J
\bsigma_\alpha\cdot\bU_k\bsigma^\prime}\right\} \nonumber
\\
\hspace*{-20mm} && \hspace*{-35mm} =\int\!\{\rmd\! P\}
\prod_\alpha P(\bsigma_\alpha) \sum_{\ell\geq 0}\frac{c^\ell}{\ell
!}e^{-c} \int\!\prod_{k=1}^\ell[\{\rmd\! P_k\} W[\{P_k\}]\rmd\bU_k
P(\bU_k)]~Z^n[\{P_1,\ldots,P_\ell\}] \nonumber
\\
\hspace*{-20mm} && \hspace*{-15mm} \times \prod_{\bsigma\in
S_d}\delta\left[ P(\bsigma)- \frac{e^{-\beta V(\bsigma)}
\prod_{k=1}^\ell \int\!\rmd\bsigma^\prime P_k(\bsigma^\prime)
e^{\beta J
\bsigma\cdot\bU_k\bsigma^\prime}}{Z[\{P_1,\ldots,P_\ell\}]}\right]
\end{eqnarray}
where
 \be
  Z[\{P_1,\ldots,P_\ell\}]=\int\!\rmd\bsigma~e^{-\beta
V(\bsigma)} \prod_{k=1}^\ell \int\!\rmd\bsigma^\prime
P_k(\bsigma^\prime) e^{\beta J \bsigma\cdot\bU_k\bsigma^\prime}
 \ee
 In the replica
limit $n\to 0$, both the term $Z^n[\{P_1,\ldots,P_k\}]$ and the
constant $C_n$ reduce to unity, and our RS order parameter
equation acquires the transparent form
\begin{eqnarray}
\hspace*{-15mm}
 W[\{P\}]&=& \sum_{\ell\geq 0} \frac{c^\ell}{\ell
!}e^{-c}\!\int\!\prod_{k\leq \ell}[\{\rmd\! P_k\}W[\{P_k\}]
\rmd\bU_k P(\bU_k)]
 \nonumber \\
 \hspace*{-15mm}
 && \times \prod_{\bsigma\in S_{d-1}}
\delta\left[ P(\bsigma)- \frac{e^{-\beta V(\bsigma)}
\prod_{k=1}^\ell \int\!\rmd\bsigma^\prime P_k(\bsigma^\prime)
e^{\beta J \bsigma\cdot\bU_k\bsigma^\prime}}
{\int\!\rmd\bsigma^\pprime e^{-\beta V(\bsigma^\pprime)}
\prod_{k=1}^\ell \int\!\rmd\bsigma^\prime P_k(\bsigma^\prime)
e^{\beta J \bsigma^\pprime\cdot\bU_k\bsigma^\prime}} \right]
\label{eq:population_dynamics}
\end{eqnarray}
The replica symmetric order parameter for finitely and randomly
connected disordered systems with continuous degrees of freedom is
thus seen to be a functional $W[\{P\}]$ on the space of
probability densities. For any specific parametrization
$P[\bsigma|\param]$ of this space, the order parameter equation
(\ref{eq:population_dynamics}) becomes an equation for the
distribution  $w(\param)$ of parameters (of which there must
generally be an infinite number):
\begin{eqnarray}
\hspace*{-15mm} w(\param)&=& \sum_{\ell\geq 0} \frac{c^\ell}{\ell
!}e^{-c}\!\int\!\prod_{k\leq \ell}[\rmd\param_k w(\param_k)
\rmd\bU_k P(\bU_k)]\int\!\{\rmd\! P\}\delta[\param-\param(\{P\})]
 \nonumber \\
 \hspace*{-15mm}
 && \times \prod_{\bsigma\in S_{d-1}}
\delta\left[ P(\bsigma)- \frac{e^{-\beta V(\bsigma)}
\prod_{k=1}^\ell \int\!\rmd\bsigma^\prime
P[\bsigma^\prime|\param_k] e^{\beta J
\bsigma\cdot\bU_k\bsigma^\prime}} {\int\!\rmd\bsigma^\pprime
e^{-\beta V(\bsigma^\pprime)} \prod_{k=1}^\ell
\int\!\rmd\bsigma^\prime P[\bsigma^\prime|\param_k] e^{\beta J
\bsigma^\pprime\cdot\bU_k\bsigma^\prime}} \right]
\label{eq:population_dynamics_specific}
\end{eqnarray}
We note that multiplication of $P[\bsigma|\param]$ by a constant
will not affect (\ref{eq:population_dynamics_specific}), so that
in our parametrizations $P[\bsigma|\param]$ we need not impose
normalization explicitly. Application of the above manipulations
to  formula (\ref{eq:RSB_f}) leads us in a similar manner to the
following expression for the RS free energy:
\begin{eqnarray}
\hspace*{-22mm} \overline{f}_{\rm RS}
  &=&
 \frac{c}{2\beta }\int\!\{\rmd\! P_1
\rmd\! P_2\}~W[\{P_1\}]W[\{P_2\}] \int\!\rmd\bU P(\bU) \log
\left[\int\!\rmd\bsigma \rmd\bsigma^\prime P_1(\bsigma)
P_2(\bsigma^\prime) e^{\beta J
\bsigma\cdot\bU\bsigma^\prime}\right] \nonumber
\\
\hspace*{-22mm} &&
 -  \frac{1}{\beta }
\sum_{\ell\geq 0}\frac{c^\ell}{\ell!} e^{-c}
\int\!\prod_{k=1}^\ell[\{\rmd\! P_k\}W[\{P_k\}]\rmd\bU_k P(\bU_k)]
\nonumber
\\
\hspace*{-22mm} && \hspace*{40mm} \times
 \log \left\{\int\!\rmd\bsigma~e^{-\beta V(\bsigma)}\prod_{k=1}^\ell \int\!\rmd\bsigma^\prime
  P_k(\bsigma^\prime)
 e^{\beta J \bsigma\cdot\bU_k \bsigma^\prime}\right\}
 \label{eq:fRS_W}
\end{eqnarray}

%%%%%%%%%%%%%%%%%%%%%%%%%%%%%%%%%%%%%%%%%%%%%%%%%%%%%%%%%%%%%%%%%
\subsection{The Ising limit}
%%%%%%%%%%%%%%%%%%%%%%%%%%%%%%%%%%%%%%%%%%%%%%%%%%%%%%%%%%%%%%%%%

As a simple consistency test of our theory, we now consider the
limit $\gamma\to\infty$ of our population dynamics equation
(\ref{eq:population_dynamics}), for the special choice $V(\bsigma)
= \gamma (\hat{\mathbf{e}} \ldotp\bsigma)^2$ where
$\hat{\mathbf{e}}$ denotes an arbitrary unit-length vector in
$\R^d$. Via a saddle point argument, we observe that the limit
$\gamma\to\infty$ restricts our spins to take one of two possible
values, viz.\@ $\bsigma=\pm \hat{\mathbf{e}}$. Hence, our
order-parameter function $P(\bsigma)$ collapses to a sum of two
delta peaks, which (with a modest amount of foresight) we choose
to parametrize by
\begin{equation}
P(\bsigma) = \frac{e^{\beta h}}{2\cosh( \beta h)} \delta(\bsigma -
\hat{\mathbf{e}}) + \frac{e^{-\beta h}}{2\cosh(\beta
h)}\delta(\bsigma + \hat{\mathbf{e}})
\end{equation}
With this representation of $P(\bsigma)$, our order parameter
functional $W[\{P\}]$ reduces to a distribution $W(h)$ of
`effective' fields $h$. Carrying out the integrations within the
$\delta$-functional in (\ref{eq:population_dynamics}) then results
in
\begin{eqnarray}
\hspace*{-20mm} W(h) &=& \sum_{\ell \geq 0}
\frac{e^{-c}c^\ell}{\ell!} \int \prod_{k \leq \ell}[\rmd h_{k}
W(h_k) \rmd\mathbf{U}_k P(\mathbf{U}_k)]
\\
\hspace*{-20mm} && \times \prod_{\bsigma \in\{-\he,\he\}} \delta
\left[\frac{e^{ \beta h (\bsigma \ldotp \he)}}{2\cosh(\beta h)} -
\frac{\prod_{k=1}^{\ell} \left[\frac{e^{\beta h_k + \beta J
\bsigma \ldotp \mathbf{U}_k \he}}{2\cosh (\beta h_k)} +
\frac{e^{-\beta h_k - \beta J \bsigma \ldotp \mathbf{U}_k
\he}}{2\cosh (\beta h_k)}\right]}{\sum_{\bsigma^\prime \in
\{-\he,\he\}} \prod_{k=1}^{\ell} \left[\frac{e^{\beta h_k + \beta
J \bsigma^\prime \ldotp \mathbf{U}_k \he}}{2\cosh (\beta h_k)} +
\frac{e^{-\beta h_k - \beta J \bsigma^\prime \ldotp \mathbf{U}_k
\he}}{2\cosh (\beta h_k)}\right]}\right] \nonumber
\end{eqnarray}
The two $\delta$-functions in this expression now effectively give
us an update relation for $h$. This leads to a population dynamics
equation, equivalent to that found in e.g.
\cite{kanter-sompo87,mezard-parisi87}:
\begin{eqnarray}
W(h) &=& \sum_{\ell \geq 0} \frac{e^{-c}c^\ell}{\ell!} \int
\prod_{k \leq \ell}[\rmd h_{k} W(h_k) \rmd \mathbf{U}_k
P(\mathbf{U}_k)]\nonumber\\ && \times\ \delta\left[h -
\frac{1}{\beta}\sum_{k=1}^{\ell} \mbox{arctanh}[\tanh(\beta h)
\tanh(\beta J \he\ldotp \mathbf{U}_k \he)]\right]
\end{eqnarray}
Choosing, for instance, the random orthogonal matrix distribution
$P({\bf U})$ to be of the simple form $P(\mathbf{U})= a P(\one) +
(1-a)P(-\one)$\footnote{Here we take the freedom to consider the
orthogonal group $O(3)$ instead of $SO(3)$, otherwise the matrix
$-\one$ would not be
 in our ensemble (defined as the orthogonal matrices with determinant one).},
with $a\in[0,1]$, leads directly to the familiar functional order
parameter equations of the $\pm J$ Ising spin-glass, with exchange
interactions distributed according to $P(J^\prime)= a
\delta[J^\prime-J] + (1-a)\delta[J^\prime+J]$.

\section{General theory for $d=2$}

\subsection{RS saddle-point equations and free energy}

For $d=2$ the set $S_{d-1}$ reduces to the unit circle, and our
spins become XY-spins, i.e. $\bsigma=(\cos\phi,\sin\phi)$ with
$\phi\in[0,2\pi]$. We may therefore also write
$V(\bsigma)=\tilde{V}(\phi)$, and find the random rotation
matrices ${\bf U}$ being simply characterized by a single angle
$\omega\in[0,2\pi]$ and an associated symmetric distribution
$P(\omega)=P(-\omega)$. As a result our equations simplify
considerably. The order parameter equation
(\ref{eq:population_dynamics}) reduces to
\begin{eqnarray}
\hspace*{-10mm}
 W[\{P\}]&=& \sum_{\ell\geq 0} \frac{c^\ell}{\ell
!}e^{-c}\!\int\!\prod_{k\leq \ell}[\{\rmd\! P_k\}W[\{P_k\}]
\rmd\omega_k P(\omega_k)]
 \nonumber \\
 \hspace*{-10mm}
 && \times \prod_{\phi}
\delta\left[ P(\phi)- \frac{e^{-\beta \tilde{V}(\phi)}
\prod_{k=1}^\ell \int\!\rmd\phi^\prime P_k(\phi^\prime) e^{\beta J
\cos(\phi-\phi^\prime-\omega_k)}} {\int\!\rmd\phi^\pprime
e^{-\beta \tilde{V}(\phi^\pprime)} \prod_{k=1}^\ell
\int\!\rmd\phi^\prime P_k(\phi^\prime) e^{\beta J
\cos(\phi^\pprime-\phi^\prime-\omega_k)}} \right]
\end{eqnarray}
In the absence of single-site potentials, i.e. for
$\tilde{V}(\phi)=0$, we get
\begin{eqnarray}
\hspace*{-10mm}
 W[\{P\}]&=& \sum_{\ell\geq 0} \frac{c^\ell}{\ell
!}e^{-c}\!\int\!\prod_{k\leq \ell}[\{\rmd\!P_k\}W[\{P_k\}]
\rmd\omega_k P(\omega_k)]
 \nonumber \\
 \hspace*{-10mm}
 && \times \prod_{\phi}
\delta\left[ P(\phi)- \frac{ \prod_{k=1}^\ell
\int\!\rmd\phi^\prime P_k(\phi^\prime) e^{\beta J
\cos(\phi-\phi^\prime-\omega_k)}} {\int\!\rmd\phi^\pprime
\prod_{k=1}^\ell \int\!\rmd\phi^\prime P_k(\phi^\prime) e^{\beta J
\cos(\phi^\pprime-\phi^\prime-\omega_k)}} \right]
\label{eq:RSspe_d2}
\end{eqnarray}
For $T=0$ this reduces even further to
\begin{eqnarray}
\hspace*{-10mm}
 W[\{P\}]&=& \sum_{\ell\geq 0} \frac{c^\ell}{\ell
!}e^{-c}\!\int\!\prod_{k\leq \ell}[\{\rmd\!P_k\}W[\{P_k\}]
\rmd\omega_k P(\omega_k)]
 \nonumber \\
 \hspace*{-10mm}
 && \times \prod_{\phi}
\delta\left[ P(\phi)- \frac{ \prod_{k=1}^\ell P_k(\phi-\omega_k)}
{\int\!\rmd\phi^\pprime \prod_{k=1}^\ell
P_k(\phi^\pprime\!-\omega_k) } \right]\label{eq:Tzero}
\end{eqnarray}
Similarly we find that for $V(\bsigma)=0$ and $d=2$ the RS free
energy per spin (\ref{eq:fRS_W}) becomes
\begin{eqnarray}
\hspace*{-25mm} \overline{f}_{\rm RS}
  &=&
 \frac{c}{2\beta }\int\!\{\rmd\!P_1
\rmd\!P_2\}~W[\{P_1\}]W[\{P_2\}] \int\!\rmd\omega P(\omega) \log
\left[\int\!\rmd\phi \rmd\phi^\prime P_1(\phi) P_2(\phi^\prime)
e^{\beta J \cos(\phi-\phi^\prime-\omega)}\right] \nonumber
\\
\hspace*{-25mm} &&
 -  \frac{1}{\beta }
\sum_{\ell\geq 0}\frac{c^\ell}{\ell!} e^{-c}
\int\!\prod_{k=1}^\ell[\{\rmd\!P_k\}W[\{P_k\}]\rmd\omega_k
P(\omega_k)]\nonumber
\\
\hspace*{-25mm}&& \hspace*{30mm}
\times
 \log \left[\int\!\rmd\phi \prod_{k=1}^\ell \int\!\rmd\phi^\prime
  P_k(\phi^\prime)
 e^{\beta J \cos(\phi-\phi^\prime-\omega_k)}\right]
 \label{eq:fRS_d2}
\end{eqnarray}
 As expected,
the paramagnetic state
$W[\{P\}]=\prod_{\phi\in[0,2\pi]}\delta[P(\phi)-(2\pi)^{-1}]$ is a
solution of (\ref{eq:RSspe_d2}) at any temperature. In this state
one finds, using (\ref{eq:physical_meaning}), that $\bra
\delta[\phi-\phi_i]\ket=(2\pi)^{-1}$ for all $i$ and all $\phi$.

Continuous bifurcations away from the paramagnetic state can be
identified via a so-called Guzai (or functional moment) expansion.
We transform $P(\phi)\to (2\pi)^{-1}+\Delta(\phi)$, with
$W[\{P\}]\to \tilde{W}[\{\Delta\}]$ and with
$\tilde{W}[\{\Delta\}]=0$ as soon as $\int_0^{2\pi}\!\rmd\phi
~\Delta(\phi)\neq 0$ (since $P(\phi)$ must remain normalized). We
may now expand our equations in powers of the functional moments
$\int\{\rmd\Delta\}\tilde{W}[\{\Delta\}]\Delta(\phi_1)\ldots
\Delta(\phi_r)$ for $r=1,2$. In doing so we will repeatedly
encounter the modified Bessel functions
\be
I_n(z)=\int_{-\pi}^\pi\!\frac{\rmd\phi}{2\pi}\cos(n\phi)e^{z\cos(\phi)}
\ee
 Close to a continuous phase transition we assume there to be a
small parameter $\epsilon$ measuring the bifurcation, such that
$\int\{\rmd\Delta\}\tilde{W}[\{\Delta\}]\Delta(\phi_1)\ldots
\Delta(\phi_r)=\order(\epsilon^r)$.

\subsection{Paramagnetic to ferromagnetic and Kosterlitz-Thouless type transitions}

In lowest order $\epsilon^1$ we have
\begin{eqnarray}
\hspace*{-25mm}
\int\!\{\rmd\Delta\}\tilde{W}[\{\Delta\}]\Delta(\phi) &=&
\frac{1}{2\pi}\sum_{\ell\geq 0} \frac{c^\ell}{\ell
!}e^{-c}\!\int\!\prod_{k\leq
\ell}[\{\rmd\Delta_k\}\tilde{W}[\{\Delta_k\}] \rmd\omega_k
P(\omega_k)]
 \nonumber \\
 \hspace*{-25mm}
 && \times\left\{
 \frac{ 1+\sum_{k=1}^\ell
\frac{ \int\!\rmd\phi^\prime \Delta_k(\phi^\prime) e^{\beta J
\cos(\phi-\phi^\prime-\omega_k)} }{ I_0(\beta J)}+\ldots} {1
+\sum_{k=1}^\ell \int\!\frac{\rmd\phi^\pprime}{2\pi}
\frac{\int\!\rmd\phi^\prime\Delta_k(\phi^\prime) e^{\beta J
\cos(\phi^\pprime-\phi^\prime-\omega_k)}}{ I_0(\beta J)} +\ldots}
-1 \right\} \nonumber
\\
 \hspace*{-25mm}
&& \hspace*{-25mm}= \frac{1}{2\pi}\sum_{\ell\geq 0}
\frac{c^\ell}{\ell !}e^{-c}\!\int\!\prod_{k\leq
\ell}[\{\rmd\Delta_k\}\tilde{W}[\{\Delta_k\}] \rmd\omega_k
P(\omega_k)]
 \nonumber \\
 \hspace*{-25mm}
 && \hspace*{-22mm}\times  \sum_{k=1}^\ell\left\{
\frac{ \int\!\rmd\phi^\prime \Delta_k(\phi^\prime) e^{\beta J
\cos(\phi-\phi^\prime-\omega_k)} }{ I_0(\beta J)} -
\int\!\frac{\rmd\phi^\pprime}{2\pi}
\frac{\int\!\rmd\phi^\prime\Delta_k(\phi^\prime) e^{\beta J
\cos(\phi^\pprime-\phi^\prime-\omega_k)}}{ I_0(\beta J)} +\ldots
\right\} \nonumber
\\
 \hspace*{-25mm}
&&\hspace*{-25mm} =\frac{c}{2\pi I_0(\beta
J)}\int\!\{\rmd\Delta\}\tilde{W}[\{\Delta\}] \rmd\omega P(\omega)
\int\!\rmd\phi^\prime \Delta(\phi^\prime) e^{\beta J
\cos(\phi-\phi^\prime-\omega)} +\order(\epsilon^2)
\end{eqnarray}
Thus, with
$\Psi(\phi)=\int\{\rmd\Delta\}\tilde{W}[\{\Delta\}]\Delta(\phi)$
we obtain the following (constrained) leading order eigenvalue
problem, which describes transitions away from the paramagnetic
state:
\begin{eqnarray}
 \Psi(\phi)=\frac{c}{2\pi I_0(\beta J)}
\int_0^{2\pi}\!\!\rmd\phi^\prime \int\! \rmd\omega~
P(\omega)e^{\beta J
\cos(\phi-\phi^\prime-\omega)}\Psi(\phi^\prime)
\\
 \int_0^{2\pi}\!\!\rmd\phi~\Psi(\phi)=0
 \end{eqnarray}
  This
problem is solved by the Fourier modes $\Psi(\phi)=\hat{\Psi}_k
e^{\rmi k\phi}$ (with integer $k\neq 0$), each of which bifurcates
at a temperature $T_k$ which is to be solved from
\be
1 =\frac{cI_k(\beta J)}{I_0(\beta J)} \int_{-\pi}^{\pi}\!\!
\rmd\omega~ P(\omega)\cos(k\omega)
 \ee
For either $\beta\to 0$ or $c\to 0$ the right-hand side would
reduce to zero, and we would find ourselves always in a
paramagnetic state. The presently studied  transition therefore
occurs at the average connectivity $c$ for which
\be
c =\min_{k>0}\left\{\frac{ I_k(\beta J)}{I_0(\beta J)}
\int_{-\pi}^{\pi}\!\! \rmd\omega~
P(\omega)\cos(k\omega)\right\}^{-1} \label{eq:PtoForSG1}
 \ee
At zero temperature we may use the property
$\lim_{z\to\infty}I_k(z)/I_0(z)=1$ to obtain $c^{-1}_{\rm
crit}=\max_{k>0}\int_{-\pi}^{\pi} \rmd\omega~
P(\omega)\cos(k\omega)\leq 1$. According to
(\ref{eq:physical_meaning}), the present type of bifurcation is
towards a state where (with $k$ denoting the critical Fourier
mode):
\be
\lim_{N\to\infty}\frac{1}{N}\sum_i\overline{ \bra
\delta[\phi-\phi_i]\ket}=\frac{1}{2\pi}\left[1+\epsilon
\cos(k\phi-\psi)+\ldots\right]
 \ee
Care is to be taken in interpreting the bifurcating state, since
in leading order one has
\begin{eqnarray}
\lim_{N\to\infty}\frac{1}{N}\sum_i\overline{ \bra\bsigma_i\ket}&=&
\lim_{N\to\infty}\frac{1}{N}\sum_i\overline{
\bra\left(\!\begin{array}{c}\cos(\phi_i)\\
\sin(\phi_i)\end{array}\!\right)\ket}\nonumber \\ &=&
\frac{\epsilon}{2\pi} \left(\!\begin{array}{c}\int_{-\pi}^\pi
\rmd\phi~ \cos(\phi)\cos(k\phi-\psi)\\ \int_{-\pi}^\pi
\rmd\phi~\sin(\phi)\cos(k\phi-\psi)\end{array}\!\right)+\ldots\nonumber
\\
&=& \frac{1}{2}\epsilon ~\delta_{k1}
 \left(\!\begin{array}{c}\cos(\psi)\\ \sin(\psi)\end{array}\!\right)+\ldots
\end{eqnarray}
We conclude that only for $k=1$ may we call the bifurcating
solution ferromagnetic (F). For $k>1$ we find a bifurcation
towards a state with no overall magnetization, but still with
measurably non-uniform overall single spin statistics. This
transition is reminiscent of a Kosterlitz-Thouless one. Thus we
have the following possible transitions
\begin{eqnarray}
 {\rm P}\to{\rm F}:~~~~~~~~
&& c =\left\{\frac{ I_1(\beta J)}{I_0(\beta J)}
\int_{-\pi}^{\pi}\!\! \rmd\omega~
P(\omega)\cos(\omega)\right\}^{-1} \label{eq:PtoF}
 \\
 {\rm KT}:~~~~~~~~
&& c = \min_{k>1}\left\{\frac{ I_k(\beta J)}{I_0(\beta J)}
\int_{-\pi}^{\pi}\!\! \rmd\omega~
P(\omega)\cos(k\omega)\right\}^{-1} \label{eq:KT}
 \end{eqnarray}
It will be shown below, however, that the KT transition is always
preceded by a spin-glass transition, and hence it is non-physical.

\subsection{Paramagnetic to spin-glass transition}

In those cases where the transition away from the paramagnetic
state is towards a new state with
$\int\{\rmd\Delta\}\tilde{W}[\{\Delta\}]\Delta(\phi)=0$ (which
would be a spin-glass state), the lowest relevant order in our
expansions is $\epsilon^2$, and we find after functional moment
expansion:
\begin{eqnarray}
\hspace*{-15mm}
\int\!\{\rmd\Delta\}\tilde{W}[\{\Delta\}]\Delta(\phi_1)\Delta(\phi_2)
&=& \frac{c}{(2\pi)^2 I_0^2(\beta J)}
\int\![\{\rmd\Delta\}\tilde{W}[\{\Delta\}] \rmd\omega P(\omega)]
 \nonumber \\
 \hspace*{-15mm}
 && \hspace*{-35mm}
 \left( \int\!\rmd\phi^\prime \Delta(\phi^\prime) e^{\beta J
\cos(\phi_1-\phi^\prime-\omega)}  \right)
 \left(\int\!\rmd\phi^\prime \Delta(\phi^\prime) e^{\beta J
\cos(\phi_2-\phi^\prime-\omega)}  \right)
 +\order(\epsilon^3)
\end{eqnarray}
Thus, with
$\Psi(\phi_1,\phi_2)=\int\{\rmd\Delta\}\tilde{W}[\{\Delta\}]\Delta(\phi_1)\Delta(\phi_2)$
we now arrive at the following (constrained) eigenvalue problem
for the P$\to$SG transition:
\begin{eqnarray}
\hspace*{-20mm} \Psi(\phi_1,\phi_2) = c
\int\!\frac{\rmd\phi_1^\prime \rmd\phi_2^\prime}{[2\pi I_0(\beta
J)]^2} \left[ \int\!\rmd\omega~ P(\omega)
  e^{\beta J
\cos(\phi_1-\phi_1^\prime-\omega)+\beta J
\cos(\phi_2-\phi_2^\prime-\omega)}\right]
\Psi(\phi_1^\prime,\phi_2^\prime)
\\
\hspace*{-20mm}
\int\!\rmd\phi_1~\Psi(\phi_1,\phi_2)=\int\!\rmd\phi_2~\Psi(\phi_1,\phi_2)=0
\end{eqnarray}
Again one finds that the Fourier modes are the relevant solutions.
Here we have $\Psi(\phi_1,\phi_2)=\hat{\Psi}_{k_1,k_2}
e^{\rmi(k_1\phi_1+k_2\phi_2)}$ (with integer $k_r\neq 0$), each of
which bifurcates at a temperature (or, equivalently, at a critical
connectivity) which is to be solved from
\be
 1  =\frac{ c I_{k_1}(\beta J) I_{k_2}(\beta J)}{I_0^2(\beta
J)}
 \int\!\rmd\omega~
P(\omega)\cos[(k_1+k_2)\omega] \ee As before the right-hand side
becomes zero for $\beta=0$ or $c\to 0$, so that the transition
occurs at
\be
 c =\min_{k_1\neq
0,~k_2\neq 0}\left\{\frac{ I_{k_1}(\beta J) I_{k_2}(\beta
J)}{I_0^2(\beta J)}
 \int\!\rmd\omega~
P(\omega)\cos[(k_1+k_2)\omega]\right\}^{-1} \ee
 Since one always has
$I_k(\beta J)\geq 0$ and $I_{-k}(z)=I_k(z)$, we find that the
required extremum occurs for $k_1=-k_2=k$, upon which one
subsequently deduces that $k=1$, so
\be
 {\rm P}\to{\rm SG}:~~~~~~~~ c =I^2_{0}(\beta J)/I^2_1(\beta
 J)
 \label{eq:PtoSG}
\ee
 This equation is obviously independent of the distribution of
chiral interactions. At zero temperature we find $c_{\rm crit}=1$.
According to (\ref{eq:physical_meaning}), the present type of
bifurcation is towards a state where
\be
\hspace*{-10mm} \lim_{N\to\infty}\frac{1}{N}\sum_i\overline{ \bra
\delta[\phi-\phi_i]\ket \bra \delta[\phi^\prime-\phi_i]\ket}=
\frac{1}{(2\pi)^{2}}\left[1\!+ \epsilon
\cos(\phi-\phi^\prime+\psi)+\ldots\right]~~~~~~~~
 \ee
 We note that
 this particular type of bifurcation
obeys  $\lim_{N\to\infty}N^{-1}\sum_i\overline{ \bra
\delta[\phi-\phi_i]\ket}=(2\pi)^{-1}$, i.e. absence of measurable
overall non-uniform spin statistics. Nevertheless,
$\lim_{N\to\infty}N^{-1}\sum_i\overline{ \bra \bsigma_i\ket^2}>0$,
so the bifurcating solution describes a spin-glass state
(SG).

\subsection{Summary of transitions away from paramagnetic state and special limits}
\label{sec:bifurcations}

We define in accordance with the results
(\ref{eq:PtoF},\ref{eq:PtoSG}):
\begin{eqnarray}
c^{-1}_{\rm F} &=& \frac{ I_1(\beta J)}{I_0(\beta J)}
\int_{-\pi}^{\pi}\! \rmd\omega~ P(\omega)\cos(\omega)
\label{eq:ccrit_F}
\\
c^{-1}_{\rm SG} &=& I^2_{1}(\beta J)/I^2_0(\beta
 J)
 \label{eq:ccrit_SG}
 \end{eqnarray}
  The physical transition\footnote{We will for now leave aside the
possibility of first order transitions.} away from the
paramagnetic state, as the connectivity is increased from $c=0$
for fixed $\beta J$, is the one with the largest value of
$c^{-1}_{\rm crit}$.
 The Kosterlitz-Thouless type transition
(\ref{eq:KT}) was given by \bd c^{-1}_{\rm KT} =\max_{k>1}\left\{
\frac{ I_k(\beta J)}{I_0(\beta J)} \int_{-\pi}^{\pi}\! \rmd\omega~
P(\omega)\cos(k\omega)\right\} \leq \frac{I_2(\beta J)}{I_0(\beta
J)} \ed It follows from the properties of the modified Bessel
functions \cite{AS} that $\frac{\rmd}{\rmd
z}[I_2(z)I_0(z)-I_1^2(z)]=\frac{1}{2}I_0(z)[I_3(z)-I_0(z)]< 0$.
Hence $I_2(z)/I_0(z)-I_1^2(z)/I_0^2(z)\leq 0$ with equality only
for $z=0$. This implies that $c^{-1}_{\rm SG}\geq c^{-1}_{\rm KT}$
so that the bifurcation (\ref{eq:KT}) is indeed unphysical.
 For small and large temperatures one obtains
the limiting behaviour
\begin{eqnarray}
\lim_{T\to 0}c^{-1}_{\rm F} = \int_{-\pi}^{\pi}\! \rmd\omega~
P(\omega)\cos(\omega) \leq 1 &~~~~~~~~&
 \lim_{T\to\infty} c^{-1}_{\rm F} =0
\\
\lim_{T\to 0}c^{-1}_{\rm SG} =1
 &~~~~~~~~&
 \lim_{T\to\infty}
c^{-1}_{\rm SG} = 0
 \end{eqnarray}
To recover known results in the $c\to\infty$ limit one must first
re-scale the bond strength $J$. In the case of an overall balance
towards (anti-)ferromagnetism, i.e. for chirality distributions
such that $\int_{-\pi}^{\pi} \rmd\omega~
P(\omega)\cos(\omega)=\order(c^0)\neq 0$, we have to re-scale
according to $J= \tilde{J}/c$. This gives upon taking
$c\to\infty$, and using $I_n(z) = (z/2)^n/n!
 +\order(z^{n+1})$:
\begin{eqnarray}
T_{\rm F} =  \frac{1}{2}\tilde{J} \int_{-\pi}^{\pi}\! \rmd\omega~
P(\omega)\cos(\omega)~~~~~~~~(J=\tilde{J}/c,~c\to\infty)
 \end{eqnarray}
Here there is no transition towards an SG state ever.
 In the absence of such a dominant balance, i.e. for distributions such that $\int_{-\pi}^{\pi}\rmd\omega~ P(\omega)\cos(\omega)= \Lambda/\sqrt{c}$,
 the appropriate re-scaling is $J=\tilde{J}/\sqrt{c}$. Now we find
\begin{eqnarray}
 T_{\rm F} = \frac{1}{2}\tilde{J}\Lambda
&~~~~~~~~& T_{\rm
SG}=\frac{1}{2}\tilde{J}~~~~~~~~(J=\tilde{J}/\sqrt{c},~c\to\infty)
 \end{eqnarray}
 Hence as $c\to\infty$ and the temperature is reduced from the paramagnetic state,
 for $\Lambda>1$ we will enter a ferromagnetic state, and
 for $\Lambda<1$ a spin-glass one.
 These results are in full agreement with those obtained earlier for fully
 connected systems, see e.g. \cite{CP}.

\section{Results for specific chirality distributions at $d=2$}

Let us now work out our transition lines
(\ref{eq:ccrit_F},\ref{eq:ccrit_SG}) for specific choices for the
(symmetric) chirality distribution $P(\omega)$. It is clear from
(\ref{eq:ccrit_SG}) that our choices will only affect the P$\to$F
transition. We note that so far we only have expressions for
transitions away from the paramagnetic state;
 we are not yet able to
determine the F$\to$SG transition (when both F and SG phases
exist) analytically, since this would require us to solve our
equations below the P$\to$F and/or P$\to$SG transition
temperatures. However we may put forward the conjecture (on the
basis of our experience with more conventional disordered spin
models, e.g. \cite{SK,RSB}),
 that, especially upon taking
RSB into account (if needed),  there will be no change of phase
type after the onset of order as the temperature is lowered from
$T=\infty$ to $T=0$. This conjecture would predict the elusive
F$\to$SG transition to be the horizontal line segment in the
$(T,c^{-1})$ phase diagram going from $T=0$ to the point where the
P$\to$F and P$\to$SG lines meet (the Parisi-Toulouse hypothesis
\cite{parisitoulouse}).

\begin{figure}[t]
\vspace*{-2mm} \hspace*{45mm} \setlength{\unitlength}{0.75mm}
\begin{picture}(200,80)
\put(10,15){\epsfysize=60\unitlength\epsfbox{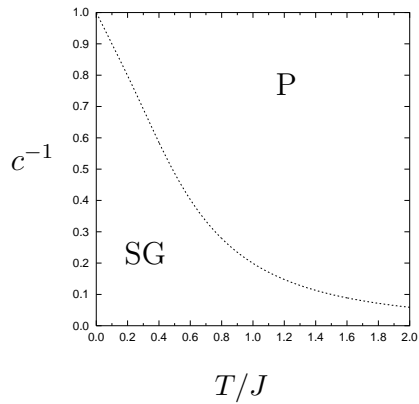}}
\put(42,6){\here{\small $T/J$}} \put(5,47){\here{$c^{-1}$}}
\put(50,60){\here{P}}
 \put(25,30){\here{SG}}
\end{picture}
\vspace*{-8mm} \caption{Predicted phase diagram for uniformly
distributed chiralities $P(\omega)=(2\pi)^{-1}$ and planar spins
($d=2$). It describing a paramagnetic (P) and a spin-glass (SG)
phase, separated by a continuous phase transition (dotted line).}
\label{fig:diagrams_uniform}
\end{figure}

\subsection{Predicted phase diagrams}

\begin{figure}[t]
\vspace*{5mm} \hspace*{5mm} \setlength{\unitlength}{0.75mm}
\begin{picture}(200,80)
\put(5,15){\epsfysize=60\unitlength\epsfbox{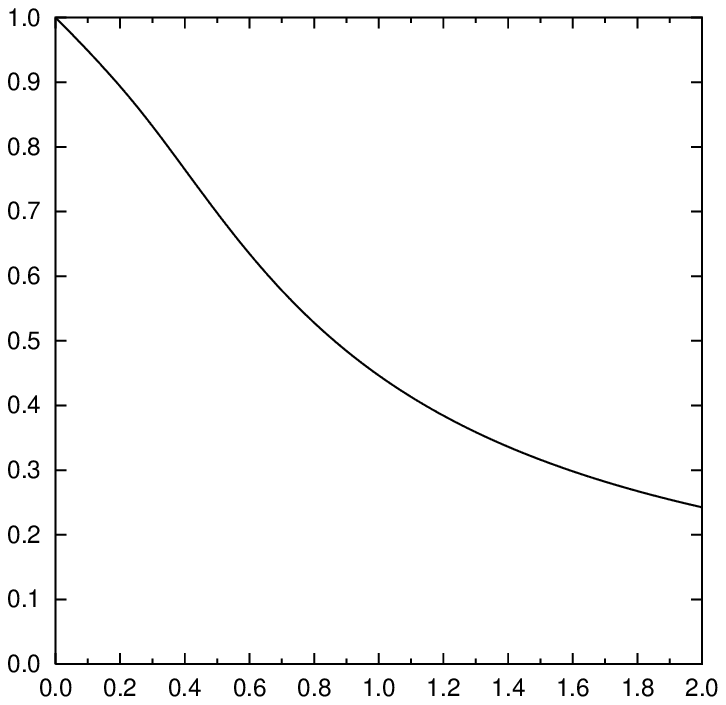}}
\put(72,15){\epsfysize=60\unitlength\epsfbox{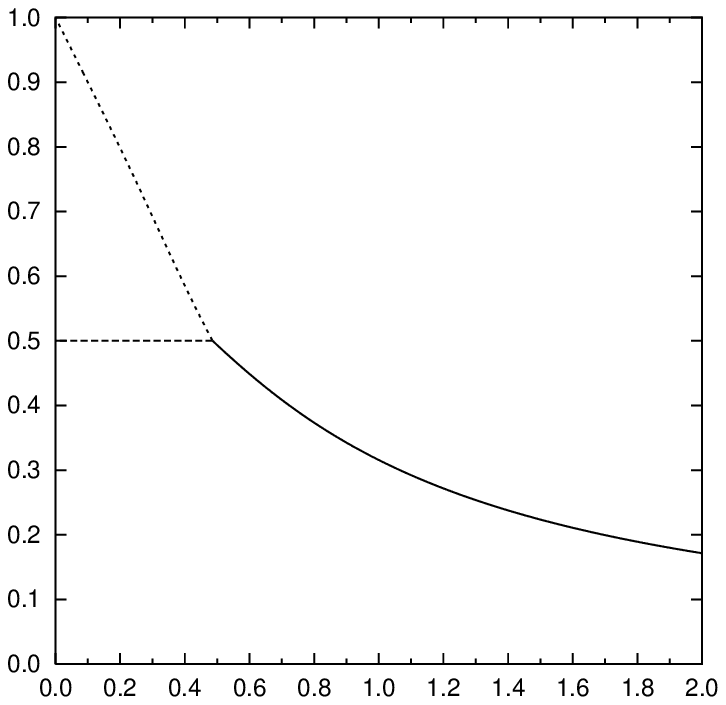}}
\put(139,15){\epsfysize=60\unitlength\epsfbox{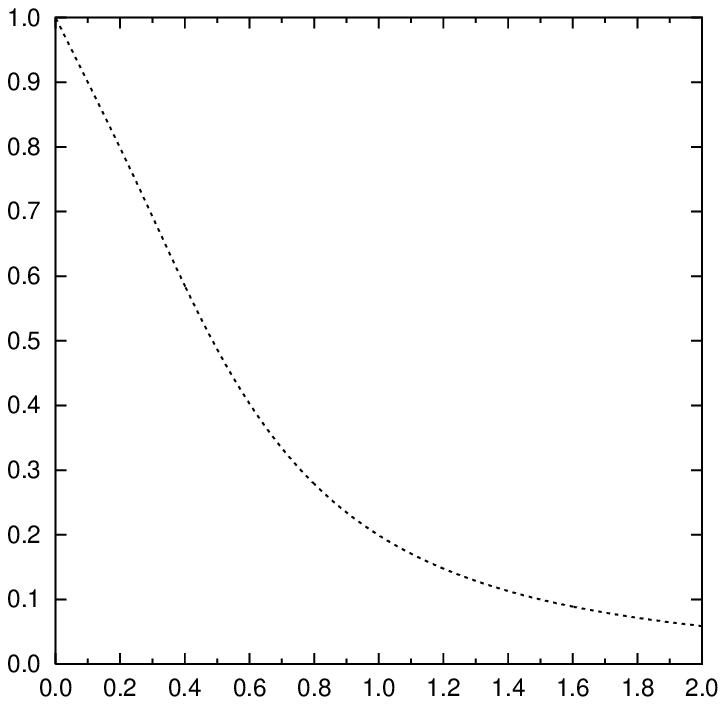}}
\put(38,6){\here{\small $T/J$}}\put(105,6){\here{\small
$T/J$}}\put(172,6){\here{\small $T/J$}}
\put(0,45){\here{$c^{-1}$}}
\put(41,80){\here{$\overline{\omega}=0$}}\put(108,80){\here{$\overline{\omega}=\frac{1}{4}\pi$}}
\put(175,80){\here{$\overline{\omega}\geq \frac{1}{2}\pi$}}
\put(48,60){\here{P}}\put(113,60){\here{P}} \put(182,60){\here{P}}
\put(21,30){\here{F}}
 \put(84,50){\here{SG}}\put(105,25){\here{F}}
\put(155,30){\here{SG}}
\end{picture}
\vspace*{-8mm} \caption{Continuous phase transitions away from the
paramagnetic (P) state for planar spins ($d=2$) and binary
chiralities
$P(\omega)=\frac{1}{2}\delta(\omega-\overline{\omega})+\frac{1}{2}\delta(\omega+\overline{\omega})$.
Solid lines: P$\to$F bifurcations.  Dotted lines: P$\to$SG
bifurcations. Note that the location of the F$\to$SG transition
(dashed) has not been calculated, but  follows from the conjecture
that on lowering temperature the nature of the ordered phase will
 remain that which emerges at the onset.} \label{fig:diagrams_binary}
\end{figure}

\begin{figure}[t]
\vspace*{5mm} \hspace*{5mm} \setlength{\unitlength}{0.75mm}
\begin{picture}(200,80)
\put(5,15){\epsfysize=60\unitlength\epsfbox{SGonly.eps}}
\put(72,15){\epsfysize=60\unitlength\epsfbox{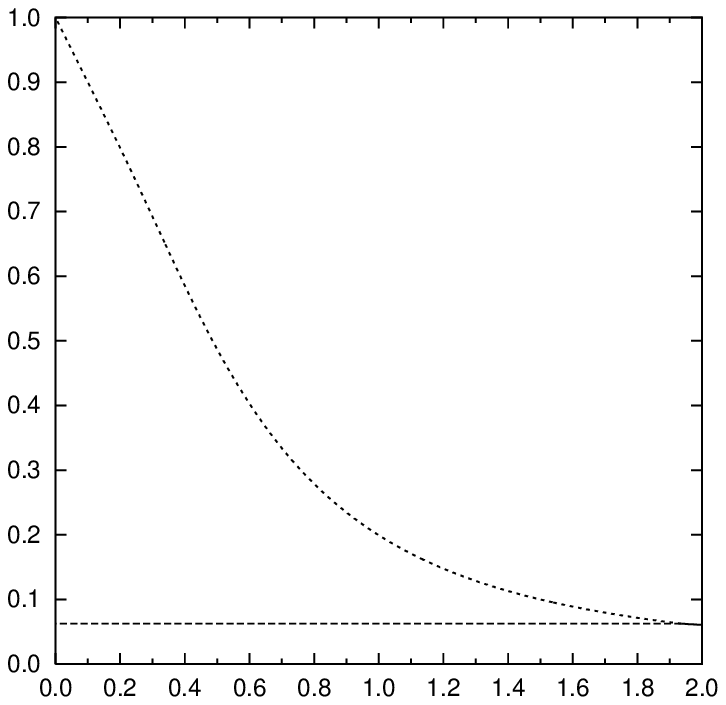}}
\put(139,15){\epsfysize=60\unitlength\epsfbox{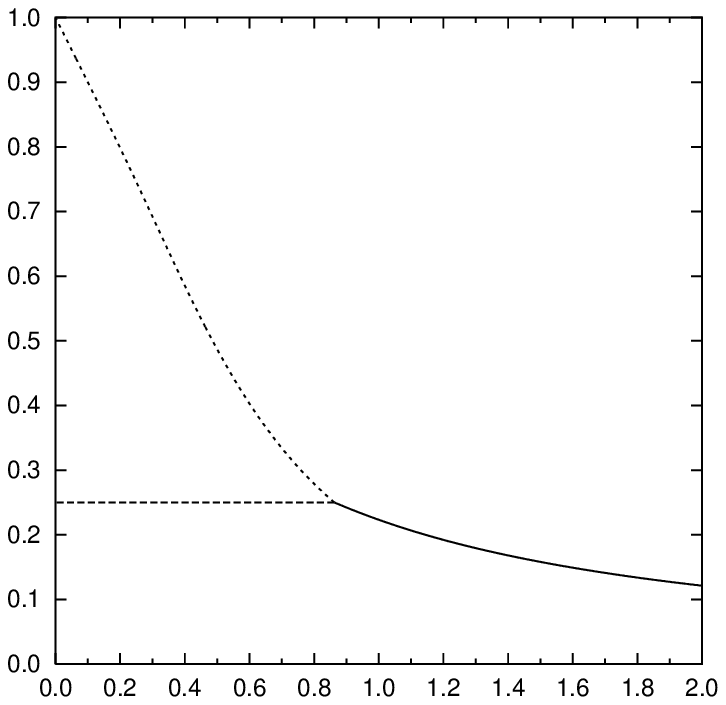}}
\put(38,6){\here{\small $T/J$}}\put(105,6){\here{\small
$T/J$}}\put(172,6){\here{\small $T/J$}}
\put(0,45){\here{$c^{-1}$}} \put(41,80){\here{$A\leq 0
$}}\put(108,80){\here{$A=\frac{1}{2}$}} \put(175,80){\here{$A=1$}}
\put(48,60){\here{P}}\put(113,60){\here{P}} \put(182,60){\here{P}}
\put(21,30){\here{SG}}
 \put(88,33){\here{SG}}
 \put(110,20){\here{F}}
 \put(153,40){\here{SG}}\put(168,24){\here{F}}
\end{picture}
\vspace*{-8mm} \caption{Continuous phase transitions away from the
paramagnetic state for planar spins ($d=2$) and resonant
chiralities $P(\omega)=[1+A\cos(\ell\phi)]/2\pi$. Solid lines:
P$\to$F bifurcations.  Dotted lines: P$\to$SG bifurcations. The
location of the F$\to$SG transition (dashed) has not been
calculated, but  follows from the conjecture that on lowering
temperature the nature of the ordered phase will
 remain that which emerges at the onset.}
\label{fig:diagrams_resonant}
\end{figure}

We first turn to uniformly distributed chiralities:
$P(\omega)=(2\pi)^{-1}$. Here we find that $c^{-1}_{\rm F}=0$ in
(\ref{eq:ccrit_F}), so for finite $c$ one only ever has the
P$\to$SG transition (\ref{eq:ccrit_SG}). The result of numerical
evaluation of the latter bifurcation line is shown in figure
\ref{fig:diagrams_uniform}.

Our second choice for the chirality statistics is the binary
distribution
$P(\omega)=\frac{1}{2}\delta(\omega-\overline{\omega})+\frac{1}{2}\delta(\omega+\overline{\omega})$,
with $\overline{\omega}\in[0,\pi]$. Here we find equations
(\ref{eq:ccrit_F},\ref{eq:ccrit_SG}) reducing to
\be
 c^{-1}_{\rm
F} = \frac{ I_1(\beta J)\cos(\overline{\omega})}{I_0(\beta J)}
~~~~~~~~  c^{-1}_{\rm SG} =\frac{I^2_{1}(\beta J)}{I^2_0(\beta
 J)}
 \label{eq:d2lines_binary}
 \ee
Now both types of transition are possible, and it will be clear
that this can result in richer phase diagrams. The P$\to$F
transitions can only occur if $\overline{\omega}\in [0,\pi/2]$;
here, as $T\to 0$ we have $0\leq \lim_{T\to 0} c^{-1}_{\rm F}
=\cos(\overline{\omega})\leq \lim_{T\to 0} c^{-1}_{\rm SG}=1$. For
$\overline{\omega}=0$ (strictly ferromagnetic forces) one will
only see a P$\to$F transition.
 As we increase $\overline{\omega}$ away from
the ferromagnetic value $\overline{\omega}=0$ we see that at the
point where  $I_1(\beta J)/I_0(\beta J)=\cos(\overline{\omega})$
the P$\to$F transition line crosses the P$\to$SG one, until at
$\overline{\omega}\geq \frac{1}{2}\pi$ the F phase has been
completely eliminated. The result of numerical evaluation of the
bifurcation lines (\ref{eq:d2lines_binary}) is shown in figure
\ref{fig:diagrams_binary}.

As a third example we inspect resonant chiralities:
$P(\omega)=[1+A\cos(\ell\phi)]/2\pi$, with $A\in[-1,1]$ and
$\ell\in\{1,2,\ldots\}$. Here we find, using $\int_{-\pi}^\pi
\rmd\omega~P(\omega)\cos(\omega)=\frac{1}{2}A\delta_{\ell 1}$,
that
 for $A\leq 0$ and also all $\ell>1$ we only have the P$\to$SG transition.
If $A>0$ and $\ell=1$, on the other hand, we may enter either the
F or the SG phase:
\be
\ell=1:~~~~~~
 c^{-1}_{\rm F} = \frac{1}{2}A \frac{ I_1(\beta
J)}{I_0(\beta J)} ~~~~~~~~ c^{-1}_{\rm SG} =\frac{I^2_{1}(\beta
J)}{I^2_0(\beta
 J)}
\ee with a possible triple point for $I_1(\beta J)/I_0(\beta
J)=\frac{1}{2}A$. The result of numerical evaluation of the
bifurcation lines is shown in figure \ref{fig:diagrams_resonant}.
Again we observe the competition between ferromagnetic and
spin-glass order.

%%%%%%%%%%%%%%%%%%%%%%%%%%%%%%%%%%%%%%%%%%%%%%%%%%%%%%%%%%%%%%%%%%%%%%%%%%%%%%%%%%%%%%%%
%%%%%%%%%%%%%%%%%%%%%%%%%%%%%%%%%%%%%%%%%%%%%%%%%%%%%%%%%%%%%%%%%%%%%%%%%%%%%%%%%%%%%%%%
\subsection{Numerical calculation of order parameters via population dynamics}
\label{sec:PDd2}
%%%%%%%%%%%%%%%%%%%%%%%%%%%%%%%%%%%%%%%%%%%%%%%%%%%%%%%%%%%%%%%%%%%%%%%%%%%%%%%%%%%%%%%%
%%%%%%%%%%%%%%%%%%%%%%%%%%%%%%%%%%%%%%%%%%%%%%%%%%%%%%%%%%%%%%%%%%%%%%%%%%%%%%%%%%%%%%%%

So far we have only shown results which did not involve solving
our order parameter equations away from bifurcation points. We now
 probe our systems further by calculating observables
(approximately) in the P and SG phases, and by comparing these
with numerical simulations. Exact execution of this work programme
 would require us to solve the functional $W[\{P\}]$ from
equation (\ref{eq:population_dynamics}). In practice one has to
resort to explicit parametrizations $P[\bsigma|\param]$ of which
the parameters $\param$ are truncated after a finite number of
components, and solve instead the truncated version of
(\ref{eq:population_dynamics_specific}). Here we choose
\be
P[\phi|\bmu]=\frac{\exp\Big[\sum_{m\geq 1}\Big( A^c_m \cos(m
\phi)+A^s_m\sin(m \phi)\Big)\Big]} {\int \!\rmd\phi^\prime~
\exp\Big[\sum_{m\geq 1}\Big( A^c_m \cos(m \phi^\prime)+A^s_m\sin(m
\phi^\prime)\Big)\Big]}
 \label{eq:parametrization}
 \ee
  with
$\bmu=(A^c_1,A^s_1,A^c_2,A^s_2,\ldots)$. Using the orthogonality
properties of $\cos(m\phi)$ and $\sin(m\phi)$ we can extract from
(\ref{eq:population_dynamics_specific}) self-consistent equations
for the measure $w(\param)$. For the simplest case
$\tilde{V}(\phi)=0$ these equations are found to take the form
\begin{eqnarray}
&& \hspace*{-25mm} w(\bmu)= \sum_{\ell\geq
0}\frac{e^{-c}c^{\ell}}{\ell!} \int [\prod_{k\leq \ell} \rmd\bmu_k
w(\bmu_k) \rmd\omega_k P(\omega_k)] \label{eq:PDmoments}
\\
&& \hspace*{-27mm} \times\! \prod_{m\geq 1}\delta\Big[A^c_m-
\sum_{k\leq \ell}\int_0^{2\pi}\!
\frac{\rmd\phi}{\pi}~\cos(m\phi)~\log\int_0^{2\pi}\!
\rmd\phi^\prime~ e^{\sum_{n>0}{ A^c_{n,k}}
\cos(n\phi^\prime)+{A^{s}_{n,k}} \sin(n\phi^\prime)+ \beta
J\cos(\phi-\phi^\prime\!-\omega_k)}\Big] \nonumber
\\
&& \hspace*{-27mm} \times\! \prod_{m\geq 1}\delta\Big[A^s_m-
\sum_{k\leq \ell}\int_0^{2\pi}\! \frac{\rmd\phi}{\pi}~\sin(m\phi)
~\log\int_0^{2\pi}\! \rmd\phi^\prime~ e^{\sum_{n>0} {A^c_{n,k}}
\cos(n\phi^\prime)+{A^s_{n,k}} \sin(n\phi^\prime)+ \beta J
\cos(\phi-\phi^\prime\!-\omega_k)}\Big] \nonumber
\end{eqnarray}
In order to solve these equations via e.g. the population dynamics
scheme \cite{mezard-parisi01}, one has to truncate the number of
coefficients in the parametrization (\ref{eq:parametrization}). In
this paper we have limited our analysis to a 2-coefficient
truncation, i.e.  $w(\bmu)\to w(A^c_1,A^s_1)$, and to populations
of size $15.10^{3}$ in the population dynamics algorithm.
Increasing the order of the parametrization by small numbers was
found to give only a modest improvement in accuracy, and it will
turn out that already at the present level  we find a good
agreement between theory and (simulation) experiments.

\begin{figure}[t]
\vspace*{-2mm} \hspace*{-3mm} \setlength{\unitlength}{0.9mm}
\begin{picture}(200,80)
\put(32,14){\includegraphics[height=60\unitlength,width=62\unitlength]{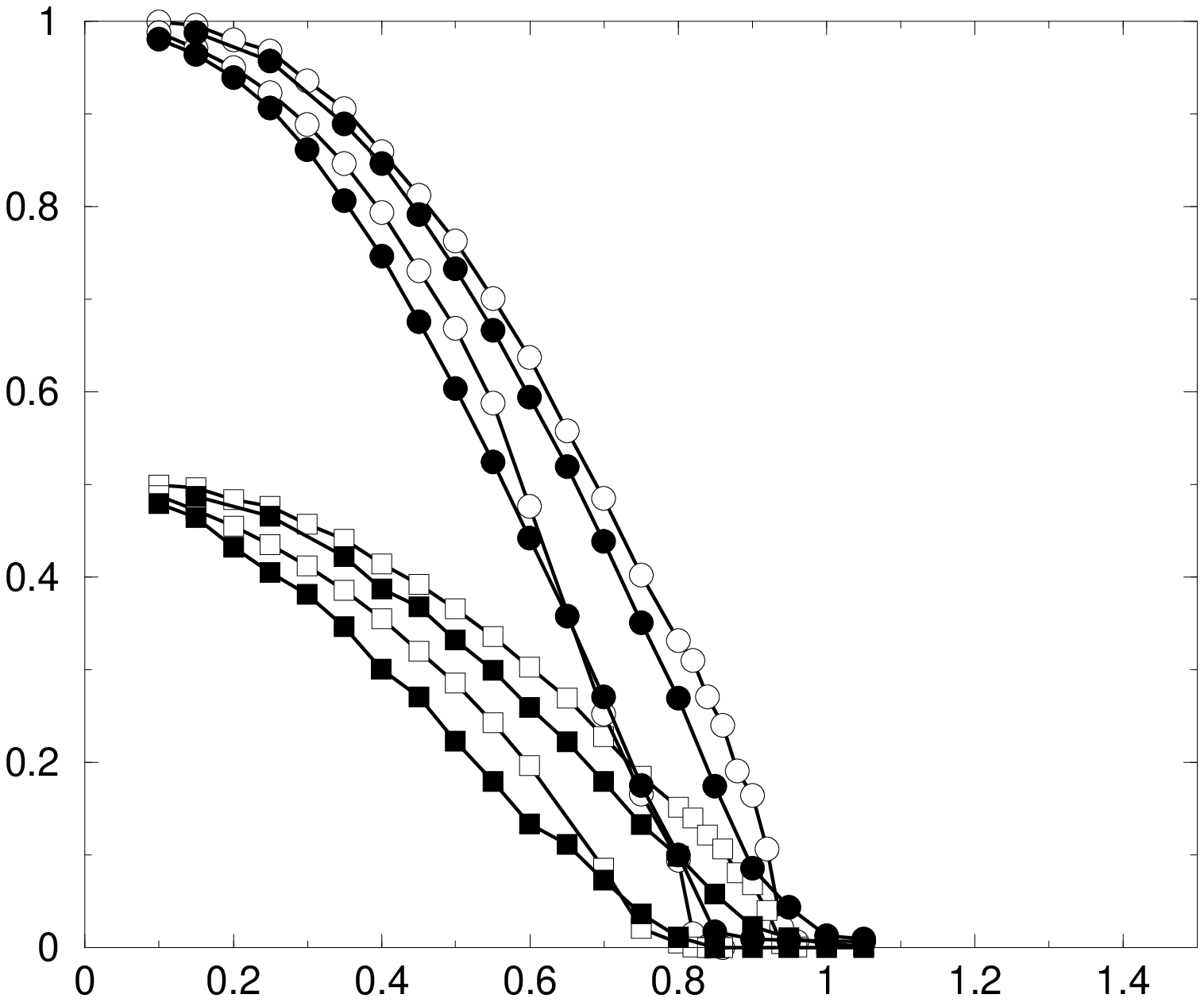}}
 \put(65,9){\here{$c^{-1}$}} \put(25,45){\here{$q,m$}}
\put(115,13){\includegraphics[height=59.6\unitlength,width=59.6\unitlength]{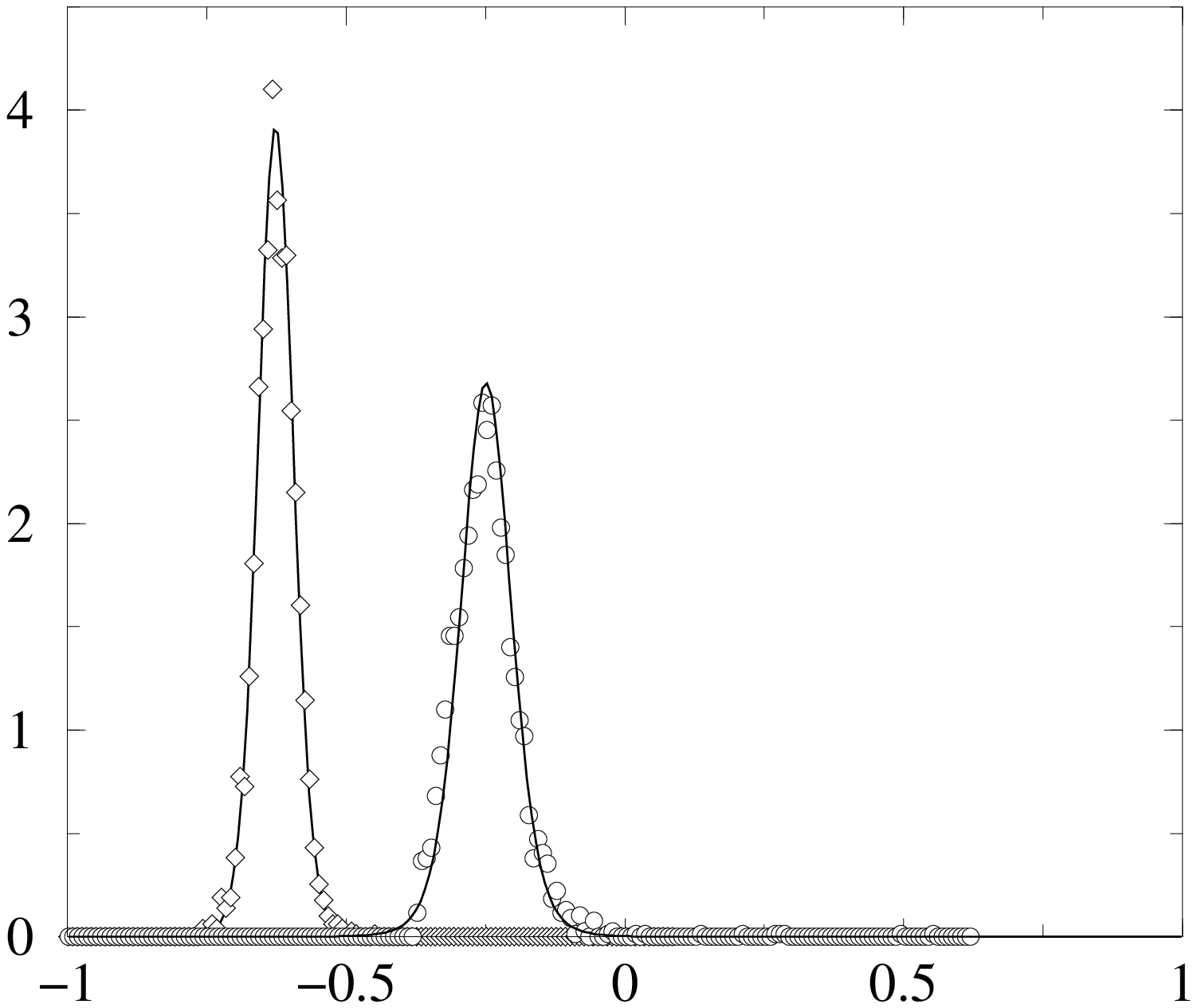}}
 \put(146,9){\here{$\phi/\pi$}} \put(106,45){\here{$P(\phi)$}}
 \put(132,62){\small $c=10$}
 \put(145,30){\small $c=5$}
\end{picture}
\vspace*{-12mm} \caption{Population dynamics calculation of
observables for strictly zero chiralities, viz.
$P(\omega)=\delta(\omega)$, and planar spins ($d=2$). Left: the
scalar observables $q=\frac{1}{2}(q_{cc}+q_{ss})$ (connected open
squares) and $m=\sqrt{m_c^2+m_s^2}$ (connected open circles) as
functions of $c^{-1}$, and for $T/J=0.1,0.3$ (upper vs. lower
solid curve). Connected full circles and squares: simulation data,
for $N=10^5$.  Right: examples at $T=0.1$ of the observed
distribution $P(\phi)=N^{-1}\sum_i\delta[\phi-\phi_i]$ in
simulations (markers), together with the corresponding theoretical
predictions (solid lines), for $c=10$ (left curves) and for $c=5$
(right curves). All observables were calculated with a population
dynamics equation (\ref{eq:PDmoments}) and a truncated
parametrization (see the main text for
 details). }
\label{fig:omegazero}
\end{figure}

The numerical simulations, of which data are shown below, were
carried out with systems of size $N=10^5$ and using the Fast
Linear Algorithm of \cite{Loison2004}. The measurements (values of
order parameters) were taken over $10^5$ iterations, following an
equilibration stage of $10^6$ iterations. The only exception is
Figure \ref{fig:omegazero}b, obtained using an Euler method with
elementary time steps of size $\Delta t= 1/2N$ and with size
$N=1000$ (after an equilibration stage of $4.10^3$ iterations per
spin).

\begin{figure}[t]
\vspace*{-2mm} \hspace*{-20mm} \setlength{\unitlength}{0.9mm}
\begin{picture}(200,80)
\put(90,15){\includegraphics[height=60\unitlength,width=62\unitlength]{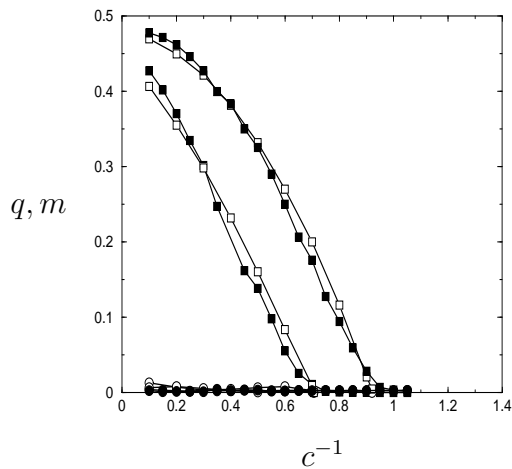}}
 \put(124,9){\here{$c^{-1}$}} \put(82,45){\here{$q,m$}}
\end{picture}
\vspace*{-12mm} \caption{Population dynamics calculation of
observables for  uniformly distributed chiralities
$P(\omega)=(2\pi)^{-1}$ and planar spins ($d=2$):
$q=\frac{1}{2}(q_{cc}+q_{ss})$ (connected open squares) and
$m=\sqrt{m_c^2+m_s^2}$ (connected open circles) as functions of
$c^{-1}$, and for $T/J=0.1,~0.3$ (upper vs. lower curves). The
observables were calculated with a population dynamics equation
(\ref{eq:PDmoments}) and a truncated parametrization (see text for
 details). Connected full squares and circles:
corresponding simulation measurements of $q$ and $m$, for systems
with $N=10^5$ spins.} \label{fig:observables_uniform}
\end{figure}

\begin{figure}[t]
\vspace*{-2mm} \hspace*{-20mm} \setlength{\unitlength}{0.9mm}
\begin{picture}(200,80)
\put(90,15){\includegraphics[height=60\unitlength,width=62\unitlength]{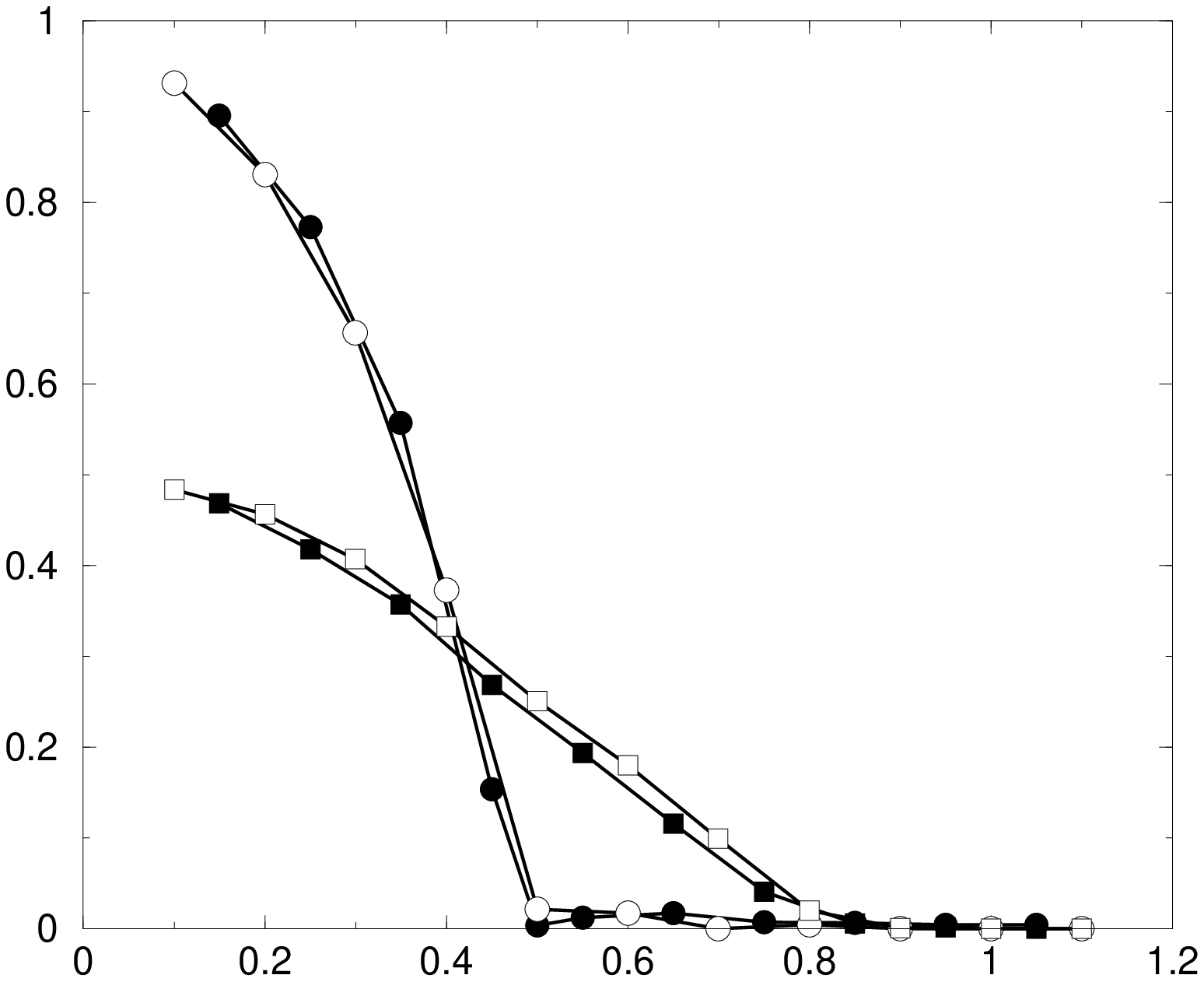}}
 \put(124,9){\here{$c^{-1}$}} \put(82,45){\here{$q,m$}}
\end{picture}
\vspace*{-12mm} \caption{Population dynamics calculation of
observables for binary distributed chiralities
$P(\omega)=\frac{1}{2}\delta[\omega+\frac{\pi}{4}]+\frac{1}{2}\delta[\omega-\frac{\pi}{4}]$
and planar spins ($d=2$): $q=\frac{1}{2}(q_{cc}+q_{ss})$
(connected open squares) and $m=\sqrt{m_c^2+m_s^2}$ (connected
open circles) as functions of $c^{-1}$, and for $T/J=0.2$ (upper
to lower solid curve).  Full markers: simulation data (full
squares: $q$; full circles: $m$), for $N=10^5$.  The observables
were calculated with a population dynamics equation
(\ref{eq:PDmoments}) and a truncated parametrization (see the main
text for
 details). }
\label{fig:omega_piover4}
\end{figure}

In testing our theory against simulation experiments we focus
mainly on the following four quantities:
\begin{eqnarray}
\hspace*{-20mm}
 m_c
 &=\lim_{N\to \infty}\frac1N \sum_{i} \overline{\Bra \cos(\phi_i)\Ket}
 &=\int\! \rmd\bmu~ w(\bmu) \int\! \rmd\phi~ P[\phi|\bmu] \cos(\phi)
\\
\hspace*{-20mm} m_s
 &= \lim_{N\to \infty}\frac1N \sum_{i}  \overline{\Bra\sin(\phi_i)\Ket}
 &= \int\! \rmd\bmu~ w(\bmu) \int\! \rmd\phi~ P[\phi|\bmu] \sin(\phi)
\\
\hspace*{-20mm} q_{cc}
 &= \lim_{N\to \infty}\frac{1}{N} \sum_{i}\overline{\Bra \cos(\phi_i)\Ket^2}
 &=\int\! \rmd\bmu~ w(\bmu) \int\! \rmd\phi \rmd\phi^\prime~ P[\phi|\bmu] P[\phi^\prime|\bmu] \cos(\phi) \cos(\phi^\prime) \label{eq:qcc}
 \\
 \hspace*{-20mm}
q_{ss}
 &= \lim_{N\to \infty}\frac{1}{N} \sum_{i}\overline{\Bra \sin(\phi_i)\Ket^2}
 &=\int\! \rmd\bmu~ w(\bmu) \int\! \rmd\phi \rmd\phi^\prime~ P[\phi|\bmu] P[\phi^\prime|\bmu] \sin(\phi) \sin(\phi^\prime) \label{eq:qss}
\end{eqnarray}
These four quantities are then compactified into the following two
scalar observables\footnote{Many equivalent choices would have
been possible.  The present definitions have the advantage that
they  will generally give $\lim_{T\to 0}q=\frac{1}{2}$ and either
$\lim_{T\to 0}m=0$ (in the SG state) or $\lim_{T\to 0}m=1$ (in the
F) state, so that we can always show both quantities together in
one plot without loss of clarity.}:
\be
q=\frac{1}{2}(q_{cc}+q_{ss}),~~~~~~~~m=\sqrt{m_c^2+m_s^2} \ee
These two quantities are sufficient for characterizing any of the
three anticipated phases $\{$P,F,SG$\}$. In the absence of
ferro-magnetism our system is invariant under global rotations,
which implies that $m_c=m_s=0$ and $q_{cc}=q_{ss}=q$. Furthermore,
in the paramagnetic state we have
$\bra\cos(\phi_i)\ket=\bra\sin(\phi_i)\ket=0$, so that
 we may write
\begin{eqnarray*}
{\rm P} &:~~~~& q=m=0
\\
{\rm F} &:~~~~& q>0,~m>0
\\
{\rm SG}&:~~~~& q>0,~m=0
\end{eqnarray*}
The results of our numerical analyses are shown in figures
\ref{fig:observables_uniform}, \ref{fig:omegazero} and
\ref{fig:omega_piover4}, and compared with simulation
measurements. We find very satisfactory agreement between theory
and simulation experiments, in spite of the combined limitations
imposed by the truncation of our parametrization in the population
dynamics analysis, the inevitable equilibration difficulties of
disordered spin systems near their transition points, and the
finite system sizes in such simulations.

 Figure \ref{fig:observables_uniform} refers to the
chirality distribution $P(\omega)=(2\pi)^{-1}$, where we should
find only the P and SG phases. This is borne out by the data: the
order parameter $m$ is indeed consistently zero, and $q$
bifurcates to a non-zero value at more or less the predicted
point. Also the locations of the transitions are as predicted by
the corresponding phase diagram, i.e Figure
\ref{fig:diagrams_uniform}.
 In figure \ref{fig:omegazero} we give data for
$P(\omega)=\delta(\omega)$. Here all interactions are strictly
ferromagnetic, leaving only the connectivity disorder, and the
only possible phases are predicted to be P and F. We do indeed
observe the predicted non-zero magnetization for small values of
$c^{-1}$, and again excellent agreement between population
dynamics  and simulations. In this figure we also show the
observed and predicted shapes\footnote{Since the system is
invariant under simultaneous rotations of all spins, the location
of the maximum of the distribution $P(\phi)$ is completely free
(only the shape carries information). To enable meaningful
comparison, one therefore first has to position the theoretical
curve such that its maximum coincides with that of the simulation
data.} of the spin angle distribution
$P(\phi)=N^{-1}\sum_{i}\delta[\phi-\phi_i]$ (these are  predicted
by the theory to equal
 $P(\phi)=\int
\rmd\bmu~ w(\bmu) P[\phi|\bmu]$), for two points in the
ferromagnetic phase.  In the SG phase such measurements tend to be
more messy and prone to finite size effects, due to the inherent
spread of the angles over the interval $[0,2\pi]$. As expected, we
see that an increase in the connectivity leads to a narrowing of
the profile of $P(\phi)$, reflecting a stronger cooperative
ordering of spin orientations. Once more the locations of the
transitions are in agreement with the phase diagram, i.e. the left
panel in Fig. \ref{fig:diagrams_binary}.
 Finally, figure
\ref{fig:omega_piover4} corresponds to the binary chirality
distribution
$P(\omega)=\frac{1}{2}\delta[\omega+\frac{\pi}{4}]+\frac{1}{2}\delta[\omega-\frac{\pi}{4}]$,
and temperature $T=0.2$. Here we  may test our assumption
regarding the location of the F$\to$SG transition line. The
prediction of the phase diagram (middle panel of Figure
\ref{fig:diagrams_binary}) is to have phase F for
$c^{-1}<\frac{1}{2}$, phase SG for $\frac{1}{2}<c^{-1}<0.798133$,
and phase P for $c^{-1}>0.798133$. The agreement between
population dynamics and simulations is once more very
satisfactory: the magnetization $m$ and the spin glass overlap $q$
indeed vanish more or less at the predicted ${\rm F}\to {\rm SG}$
and ${\rm SG}\to{\rm P}$ transition points.

  The good agreement between the truncated population dynamics results
 and our numerical simulations underlines the correctness of (i)
the RS order parameter equations themselves, (ii) the functional
moment expansion used  to locate the phase transitions P$\to$F and
P$\to$SG, and (iii) the assumed validity in the present models of
the Parisi-Toulouse hypothesis regarding the location of the
F$\to$SG transition.

\section{General theory for $d>2$}

We next try to generalize the theoretical results obtained for
$d=2$ to $d>2$. We will define the short-hand
$|S_d|=\int_{S_d}\!\rmd\bsigma$. As expected, the paramagnetic
solution of equation (\ref{eq:population_dynamics}) for
$\bsigma\in S_{d-1}$ and $V(\bsigma)=0$, is now seen to be
$P(\bsigma)=|S_{d-1}|^{-1}$ (as before, it is a solution for all
$T$). Our analysis will involve generalizations of the modified
Bessel functions, such as
$I_{0,d}(z)=|S_{d-1}|^{-1}\int_{S_{d-1}}\!d\bsigma~ e^{z\sigma_1}$
(so that $I_{0,2}(z)=I_0(z)$).

\subsection{Guzai expansion for $d>2$}

 To find bifurcations away from the paramagnetic state we
put $P(\bsigma)\to |S_{d-1}|^{-1}+\Delta(\bsigma)$, with
$W[\{P\}]\to \tilde{W}[\{\Delta\}]$ and $\tilde{W}[\{\Delta\}]=0$
as soon as $\int_{S_{d-1}}\!\rmd\bsigma ~\Delta(\bsigma)\neq 0$ so
that all probability densities are normalized, and we inspect the
lowest order functional moments
$\int\{\rmd\Delta\}\tilde{W}[\{\Delta\}]\Delta(\bsigma)$ and
$\int\{\rmd\Delta\}\tilde{W}[\{\Delta\}]\Delta(\bsigma^1)\Delta(\bsigma^2)$.
 Close to a continuous transition we assume there to be a
small parameter $\epsilon$ such that
$\int\{\rmd\Delta\}\tilde{W}[\{\Delta\}]\Delta(\bsigma^1)\ldots
\Delta(\bsigma^r)=\order(\epsilon^r)$, as before. We note that
\begin{eqnarray}
\hspace*{-25mm} \prod_{k=1}^\ell
\int_{S_{d-1}}\!\!\!\rmd\bsigma^\prime
\left[\frac{1}{|S_{d-1}|}+\Delta_k(\bsigma^\prime)\right] e^{\beta
J \bsigma\cdot\bU_k\bsigma^\prime}
 &=&  [I_{0,d}(\beta J)]^\ell\prod_{k=1}^\ell\left[
 1+
 \frac{\int_{S_{d-1}}\!\!\rmd\bsigma^\prime
\Delta_k(\bsigma^\prime) e^{\beta J
\bsigma\cdot\bU_k\bsigma^\prime}}
 {I_{0,d}(\beta J)}
 \right]\nonumber\hspace*{-10mm}
 \\
 \hspace*{-25mm}
&& \hspace*{-55mm}= [I_{0,d}(\beta J)]^\ell\left[
 1+\sum_{k\leq \ell}
 \frac{\int_{S_{d-1}}\!\rmd\bsigma^\prime
\Delta_k(\bsigma^\prime) e^{\beta J
\bsigma\cdot\bU_k\bsigma^\prime}}
 {I_{0,d}(\beta J)}
 \right.\nonumber
 \\
 \hspace*{-25mm}&&\hspace*{-50mm}
 \left.
+\frac{1}{2}\sum_{k\neq k^\prime}^\ell
 \frac{\int_{S_{d-1}}\!\!\rmd\bsigma^\prime
\Delta_k(\bsigma^\prime) e^{\beta J
\bsigma\cdot\bU_k\bsigma^\prime}}
 {I_{0,d}(\beta J)}
  \frac{\int_{S_{d-1}}\!\!\rmd\bsigma^\prime
\Delta_{k^\prime}(\bsigma^\prime) e^{\beta J
\bsigma\cdot\bU_{k^\prime}\bsigma^\prime}}
 {I_{0,d}(\beta J)}
+\order(\epsilon^3)
 \right]\nonumber\\
 \hspace*{-25mm}&&
\end{eqnarray}
Integration of this expression over $\bsigma\in S_{d-1}$ would
eliminate the $\order(\epsilon)$ term, due to the constraint
$\int_{S_{d-1}}\!\rmd\bsigma ~\Delta(\bsigma)= 0$. We may now
expand the right-hand side of (\ref{eq:population_dynamics}), and
find
\begin{eqnarray}
\hspace*{-25mm}
 \tilde{W}[\{\Delta\}]&=& \sum_{\ell\geq 0} \frac{c^\ell}{\ell
!}e^{-c}\!\int\!\prod_{k\leq
\ell}[\{\rmd\Delta_k\}\tilde{W}[\{\Delta_k\}] \rmd\bU_k P(\bU_k)]
 \prod_{\bsigma\in S_{d-1}}
\delta\left[\room \Delta(\bsigma) +\order(\epsilon^3) \right.
\nonumber \\
 \hspace*{-25mm}
&& \left.- \frac{1}{|S_{d-1}|}
 \sum_{k\leq \ell}
 \frac{\int_{S_{d-1}}\!\rmd\bsigma^\prime
\Delta_k(\bsigma^\prime) e^{\beta J
\bsigma\cdot\bU_k\bsigma^\prime}}
 {I_{0,d}(\beta J)}
\nonumber \right.
\\
\hspace*{-25mm} && \left. \hspace*{0mm}
+\frac{1}{2|S_{d-1}|}\sum_{k\neq k^\prime}^\ell\left\{
 \frac{\int_{S_{d-1}}\!\rmd\bsigma^\prime
\Delta_k(\bsigma^\prime) e^{\beta J
\bsigma\cdot\bU_k\bsigma^\prime}}
 {I_{0,d}(\beta J)}
  \frac{\int_{S_{d-1}}\!\rmd\bsigma^\prime
\Delta_{k^\prime}(\bsigma^\prime) e^{\beta J
\bsigma\cdot\bU_{k^\prime}\bsigma^\prime}}
 {I_{0,d}(\beta J)}
 \right.\right.
 \nonumber
 \\
 \hspace*{-25mm}
 && \left.\left. \hspace*{0mm}
 -\int\!\frac{d\bsigma^{\pprime}}{|S_{d-1}|}
 \frac{\int_{S_{d-1}}\!\rmd\bsigma^\prime
\Delta_k(\bsigma^\prime) e^{\beta J
\bsigma^\pprime\cdot\bU_k\bsigma^\prime}}
 {I_{0,d}(\beta J)}
  \frac{\int_{S_{d-1}}\!\rmd\bsigma^\prime
\Delta_{k^\prime}(\bsigma^\prime) e^{\beta J
\bsigma^\pprime\cdot\bU_{k^\prime}\bsigma^\prime}}
 {I_{0,d}(\beta J)}\right\}
 \right]
\end{eqnarray}
From this follow the relevant functional moment identities. We
define
$\Psi(\bsigma)=\int\{\rmd\Delta\}\tilde{W}[\{\Delta\}]\Delta(\bsigma)$
and
$\Psi(\bsigma^1,\bsigma^2)=\int\{\rmd\Delta\}\tilde{W}[\{\Delta\}]\Delta(\bsigma^1)\Delta(\bsigma^2)$.
In lowest order $\epsilon$  one now finds the condition for a
continuous P$\to$F transition by solving the (constrained)
eigenvalue problem
\begin{eqnarray}
\Psi(\bsigma) &=& \frac{c} {I_{0,d}(\beta J)}
 \int_{S_{d-1}}\frac{\!\rmd\bsigma^\prime}{|S_{d-1}|}
\Psi(\bsigma^\prime) \int\!\rmd\bU~ P(\bU) e^{\beta J
\bsigma\cdot\bU\bsigma^\prime} \label{eq:general_PtoFa}
\\ && \hspace*{-16mm}
\int_{S_{d-1}}\!\rmd\bsigma~\Psi(\bsigma)=0
\label{eq:general_PtoFb}
\end{eqnarray}
If, on the other hand, the first order to bifurcate is
$\epsilon^2$, we find a P$\to$SG transition, marked by non-trivial
solutions of the (constrained) eigenvalue problem
\begin{eqnarray}
\hspace*{-15mm}
 \Psi(\bsigma^1\!,\bsigma^2)&=&\frac{c}{I^2_{0,d}(\beta J)} \int_{S_{d-1}}\!\frac{\rmd\btau^1 \rmd\btau^2}{|S_{d-1}|^2}
\Psi(\btau^1\!,\btau^2) \int\!\rmd\bU~ P(\bU) e^{\beta J(
\bsigma^1\cdot\bU\btau^1+\bsigma^2\cdot\bU\btau^2)}
\label{eq:general_PtoSGa}
\\
&&\hspace*{-24mm}\int_{S_{d-1}}\!\rmd\bsigma^1~\Psi(\bsigma^1,\bsigma^2)=\int_{S_{d-1}}\!\rmd\bsigma^2~\Psi(\bsigma^1,\bsigma^2)=0
\label{eq:general_PtoSGb}
\end{eqnarray}
For $d=2$ these expressions
(\ref{eq:general_PtoFa},\ref{eq:general_PtoFb},\ref{eq:general_PtoSGa},\ref{eq:general_PtoSGb})
 reduce to those calculated earlier for XY spins, as they should.

\subsection{Analysis of the bifurcation conditions}

First we turn to the P$\to$F transition
(\ref{eq:general_PtoFa},\ref{eq:general_PtoFb}). It will turn out
advantageous to define two commuting linear operators
$K:S_{d-1}\to S_{d-1}$ and $L:S_{d-1}\to S_{d-1}$ as follows
 \begin{eqnarray}
  (K\Psi)(\bsigma)&=&\int_{S_{d-1}}\!\!\frac{\rmd\btau}{|S_{d-1}|}e^{\beta J \bsigma\cdot\btau} \Psi(\btau)
\label{eq:K_operator}
\\
 (L\Psi)(\bsigma)&=&\int\!\rmd\bU~
P(\bU)\Psi(\bU^\dag\bsigma) \label{eq:L_operator}
 \end{eqnarray}
 These definitions allow us to write
the eigenvalue problem
 (\ref{eq:general_PtoFa},\ref{eq:general_PtoFb})
 as
$KL\Psi=c^{-1}I_{0,d}(\beta J) \Psi$, with constraint
$\int_{S_{d-1}}\!\rmd\bsigma~\Psi(\bsigma)=0$.  We can re-write
the action of $L$ as
$(L\Psi)(\bsigma)=\int\!\rmd\bsigma^\prime~L(\bsigma,\bsigma^\prime)\Psi(\bsigma^\prime)$
with $L(\bsigma,\bsigma^\prime)= \int\!\rmd\bU~
P(\bU)\delta[\bsigma^\prime-\bU^\dag\bsigma]= \int\!\rmd\bU~
P(\bU)\delta[\bsigma-\bU\bsigma^\prime]$. The kernel
$L(\bsigma,\bsigma^\prime)$ then represents the probability that a
point $\bsigma^\prime\in S_{d-1}$ will be mapped onto $\bsigma\in
S_{d-1}$ by an orthogonal matrix from the ensemble $P(\bU)$. Both
operators $K$ and $L$ are symmetric, as a consequence of
$P(\bU)=P(\bU^\dag)$, hence we may restrict ourselves to finding
{\em simultaneous} eigenfunctions of $K$ and $L$.

It turns out that  a similar strategy can be followed for the
P$\to$SG transition
(\ref{eq:general_PtoSGa},\ref{eq:general_PtoSGb}). Here we have to
define the commuting linear operators $K:S_d\otimes S_d\to
S_d\otimes S_d$ and $L:S_d\otimes S_d\to S_d\otimes S_d$ as
 \begin{eqnarray}
  (K\Psi)(\bsigma^1,\bsigma^2)&=&\int_{S_{d-1}}\!\!\frac{\rmd\btau^1\rmd\btau^2}{|S_{d-1}|^2}e^{\beta J (\bsigma^1\cdot\btau^1+\bsigma^2\cdot\btau^2)}
  \Psi(\btau^1,\btau^2)
  \label{eq:K_operator_again}
 \\
 (L\Psi)(\bsigma^1,\bsigma^2)&=&\int\!\rmd\bU~
P(\bU)\Psi(\bU^\dag\bsigma^1,\bU^\dag\bsigma^2)
\label{eq:L_operator_again}
 \end{eqnarray}
The eigenvalue problem
 (\ref{eq:general_PtoSGa},\ref{eq:general_PtoSGb}) can now be written
 as
$KL\Psi=c^{-1}I^2_{0,d}(\beta J) \Psi$, with constraints
$\int_{S_{d-1}}\!\rmd\bsigma^1\Psi(\bsigma^1,\bsigma^2)=\int_{S_{d-1}}\!\rmd\bsigma^2\Psi(\bsigma^1,\bsigma^2)=0$.
Here we can re-write the action of $L$ as
$(L\Psi)(\bsigma^1,\bsigma^2)=\int\!\rmd\btau^1\rmd\btau^2~L(\bsigma^1,\bsigma^2,\btau^1,\btau^2)\Psi(\btau^1,\btau^2)$
with $L(\bsigma^1,\bsigma^2,\btau^1,\btau^2)= \int\!\rmd\bU~
P(\bU)\delta[\bsigma^1-\bU\btau^1]\delta[\bsigma^2-\bU\btau^2]$.
The kernel $L(\bsigma^1,\bsigma^2,\btau^1,\btau^2)$ now represents
the probability that a {\em pair} of points $(\btau^1,\btau^2)\in
S_{d-1}\otimes S_{d-1}$ will be mapped onto the pair
$(\bsigma^1,\bsigma^2)\in S_{d-1}\otimes S_{d-1}$ by an orthogonal
matrix from the ensemble $P(\bU)$. Both $K$ and $L$ are symmetric,
as a consequence of $P(\bU)=P(\bU^\dag)$, hence we may once more
restrict ourselves to finding simultaneous eigenfunctions of $K$
and $L$ individually.\vsp

At this stage it would appear appropriate to make a specific
choice for the ensemble $P(\bU)$, for which we seek a controlled
interpolation between having ferromagnetic and completely random
chiral interactions. We may define this choice in terms of the
above probability density $L(\bsigma^1,\bsigma^2,\btau^1,\btau^2)$
(from which the earlier kernel $L(\bsigma,\bsigma^\prime)$ follows
by integration), for which we now choose the linear combination
\begin{eqnarray}
L(\bsigma^1,\bsigma^2,\btau^1,\btau^2)&=&\epsilon
\delta[\bsigma^1-\btau^1]\delta[\bsigma^2-\btau^2]\nonumber
\\
&&\hspace*{-15mm}+
(1-\epsilon)\frac{\delta[|\bsigma^1|-1]\delta[|\bsigma^2|-1]
\delta[\bsigma^1\!\cdot\bsigma^2-\btau^1\!\cdot\btau^2]}{\int_{S_{d-1}}\!\rmd\bx
\rmd\by~\delta[|\bx|-1]\delta[|\by|-1]
\delta[\bx\cdot\by-\btau^1\cdot\btau^2]} \label{eq:ensemble_d3}
\end{eqnarray}
In the non-ferromagnetic part of this measure, i.e. in the term
proportional to $(1-\epsilon)$, we assign a uniform probability
density to all combined image pairs $\{\bsigma^1,\bsigma^2\}$ of
the vectors $\{\btau^1,\btau^2\}$ which preserve the inner
products under the action of the random orthogonal matrices $\bU$.
From definition (\ref{eq:ensemble_d3})  it then follows
automatically upon integration over $\bsigma^2$ that
\begin{eqnarray}
L(\bsigma,\btau)&=&\epsilon \delta[\bsigma-\btau]+
(1-\epsilon)\frac{\delta[|\bsigma|-1]}{\int_{S_{d-1}}\!\rmd\bx
~\delta[|\bx|-1]} \label{eq:ensemble_d3_simpler}
\end{eqnarray}
 For $\epsilon=1$ we return to a strictly ferromagnetic system;
for $\epsilon=0$ we have fully and homogeneously distributed
random chiral interactions.

The advantage of our choice (\ref{eq:ensemble_d3}) is that it
allows us to diagonalize both kernels
(\ref{eq:K_operator_again},\ref{eq:L_operator_again})
analytically. Working out the eigenvalue problem
$\int_{S_{d-1}}\!\rmd\bsigma^\prime~L(\bsigma,\bsigma^\prime)\Psi(\bsigma^\prime)=\lambda\Psi(\bsigma)$
shows that there are only two simple eigenspaces:
\begin{eqnarray}
\int_{S_{d-1}}\!\rmd\bsigma~\Psi(\bsigma)\neq 0:~~ \lambda=1
~~~~~{\rm and}~~~~~ \int_{S_{d-1}}\!\rmd\bsigma~\Psi(\bsigma)= 0:
~~ \lambda=\epsilon
\end{eqnarray}
The first eigenspace is forbidden by constraint
(\ref{eq:general_PtoFb}), so we may simply replace $L\to
\epsilon\one$ in the eigenvalue problem for the P$\to$F
transition. Working out the P$\to$SG eigenvalue problem
$\int_{S_{d-1}}\!\rmd\btau^1 \rmd\btau^2
~L(\bsigma^1,\bsigma^2,\btau^1,\btau^2)\Psi(\btau^1,\btau^2)=\lambda\Psi(\bsigma^1,\bsigma^2)$
leads to the following eigenspaces:
\begin{eqnarray}
\hspace*{-10mm} \int_{S_{d-1}}\!\! \rmd\bsigma^1
\rmd\bsigma^2~\delta[\bsigma^1\!\cdot\bsigma^2\!-u]\Psi(\bsigma^1\!,\bsigma^2)=0,~~{\rm
for~all}~ u\in[-1,1]: &~~~& \lambda=\epsilon
\label{eq:PSG_eveps}\\ \hspace*{-10mm}
\Psi(\bsigma^1,\bsigma^2)=\psi(\bsigma^1\!\cdot\bsigma^2),~~{\rm
for~all}~~\bsigma^1,\bsigma^2\in S_{d-1}: &~~~& \lambda=1
\label{eq:PSG_ev1}
\end{eqnarray}
It is clear that the $\lambda=1$ eigenspace is perfectly
compatible with the constraints (\ref{eq:general_PtoSGb}), which
would e.g. be satisfied by any anti-symmetric function $\psi(u)$
in (\ref{eq:PSG_ev1}).
Having solved the eigenvalue problem for the operators $L$ for the
choice of ensemble  (\ref{eq:ensemble_d3}), we may concentrate on
the following reduced eigenvalue problems from which to extract
the phase transitions away from the paramagnetic state:
\begin{eqnarray}
\hspace*{-15mm} {\rm P}\to{\rm F}: &~~~&
\int_{S_{d-1}}\!\frac{\rmd\btau}{|S_{d-1}|}e^{\beta J
\bsigma\cdot\btau} \Psi(\btau)=\frac{I_{0,d}(\beta J)}{\epsilon
c}\Psi(\bsigma) \label{eq:PF}
\\ \hspace*{-10mm}  && {\rm
constraint}:~~~\int_{S_{d-1}}\!\rmd\bsigma~\Psi(\bsigma)=0\nonumber
\\
\hspace*{-15mm} {\rm P}\to{\rm SG}_a: &~~~&
  \int_{S_{d-1}}\!\!\frac{\rmd\btau^1\rmd\btau^2}{|S_{d-1}|^2}e^{\beta J (\bsigma^1\cdot\btau^1+\bsigma^2\cdot\btau^2)}
  \Psi(\btau^1,\btau^2)=\frac{I^2_{0,d}(\beta J)}{\epsilon c}
  \Psi(\bsigma^1,\bsigma^2)
\\
\hspace*{-15mm}
 && \int_{S_{d-1}}\!\! \rmd\bsigma^1
\rmd\bsigma^2~\delta[\bsigma^1\!\cdot\bsigma^2\!-u]\Psi(\bsigma^1\!,\bsigma^2)=0,~~{\rm
for~all}~ u\in[-1,1] \nonumber
  \\
  \hspace*{-15mm}
   && {\rm
constraints}:~~~\int_{S_{d-1}}\!\rmd\bsigma^1\Psi(\bsigma^1\!,\bsigma^2)=\int_{S_{d-1}}\!\rmd\bsigma^2\Psi(\bsigma^1\!,\bsigma^2)=0
\nonumber
\\
\hspace*{-15mm} {\rm P}\to{\rm SG}_b: &~~~&
  \int_{S_{d-1}}\!\!\frac{\rmd\btau^1\rmd\btau^2}{|S_{d-1}|^2}e^{\beta J (\bsigma^1\cdot\btau^1+\bsigma^2\cdot\btau^2)}
  \psi(\btau^1\!\cdot\btau^2)=\frac{I^2_{0,d}(\beta J)}{c} \psi(\bsigma^1\!\cdot\bsigma^2)
  \label{eq:PSGb}
  \\
  \hspace*{-15mm}  && {\rm
constraint}:~~~\int_{S_{d-1}}\!\rmd\bsigma~\psi(\sigma_1)=0
\nonumber
\end{eqnarray}
Although they cannot formally be ruled out, we will henceforth
disregard the P$\to$SG$_a$ transitions, since they are less likely
to correspond to the largest eigenvalue in view of the extra
factor $\epsilon$ involved, and because in addition the associated
constraints (infinite in number) would appear to severely limit
the space of allowed functions\footnote{There are further reasons
to reduce the likelihood of the P$\to$SG$_a$ transition being
physical. For instance, for $d=2$ the eigenfunctions can only
depend on the inner product between the two vectors involved,
hence here one simply cannot satisfy the constraints of the
P$\to$SG$_a$ bifurcation.}.

\subsection{Explicit results for $d=3$}

We finally work out our previous general equations for the value
$d=3$, where we may turn to polar coordinates and write our
integration variables as
$\bsigma=(\sin(\theta)\cos(\phi),\sin(\theta)\sin(\phi),\cos(\theta))$ with
$\phi\in[-\pi,\pi]$ and $\theta\in[0,\pi]$.
We note that $I_{0,3}(x) = \sinh(x)/x$. Insertion of the polar
coordinate representation in our general eigenvalue equations
shows that for $d=3$ the second order P$\to$F transition is to be
solved from
\begin{eqnarray*}
\hspace*{-20mm}
\int_0^\pi\!\frac{\rmd\theta^\prime\sin(\theta^\prime)}{2}\int_{-\pi}^\pi\!\frac{\rmd\phi^\prime}{2\pi}~
e^{\beta
J[\sin(\theta)\sin(\theta^\prime)\cos(\phi^\prime)+\cos(\theta)\cos(\theta^\prime)]}
\Psi(\theta^\prime,\phi+\phi^\prime)=\frac{\sinh(\beta
J)}{\epsilon \beta J c}\Psi(\theta,\phi)
\end{eqnarray*}
We observe that the solutions of this equation are of the form
$\Psi(\theta,\phi)=\rho(\cos(\theta))e^{ik\phi}$ with $k\in\N$
(similar to e.g. the spherical harmonics
$Y_{\ell,m}(\theta,\phi)$, but with, as we will find below, a
different dependence on $\theta$), and with the function $\rho$ to
be solved from
 \begin{eqnarray}
\int_{-1}^1\!\frac{\rmd y}{2}~ I_{k}(\beta
J\sqrt{1-x^2}\sqrt{1-y^2}) e^{\beta J xy}\rho(y)
=\frac{\sinh(\beta J)}{\epsilon \beta J c}\rho(x)\\{\rm
constraint:~~~either}~~~k\neq 0~~~{\rm or}~~~ \int_{-1}^1\!\rmd
x~\rho(x)=0 \label{eq:problem_for_rho}
\end{eqnarray}
The kernel in (\ref{eq:problem_for_rho}) of which we seek the
eigenfunctions is invariant under parity transformation, so we may
restrict ourselves to eigenfunctions $\rho(x)$ which are either
symmetric, $\rho_+(x)$, or anti-symmetric, $\rho_-(x)$. Upon
implementing these restrictions we find
\begin{eqnarray}
\hspace*{-15mm}
 \rho_{+}(x):
&~~~&\int_{0}^1\!\rmd y~ I_{k}(\beta J\sqrt{1-x^2}\sqrt{1-y^2})
\cosh(\beta J xy)\rho(y) =\frac{\sinh(\beta J)}{\epsilon \beta J
c}\rho(x)
\\
\hspace*{-15mm}
 \rho_{-}(x):
 &~~~&
 \int_{0}^1\!\rmd y~ I_{k}(\beta J\sqrt{1-x^2}\sqrt{1-y^2} )
\sinh(\beta J xy)\rho(y) =\frac{\sinh(\beta J)}{\epsilon \beta J
c}\rho(x) \label{eq:PtoFagain}
\end{eqnarray}
It should be expected that, as for $d=2$, the physical (i.e.
largest) eigenvalue is the one with the lowest allowed value of
$k$, which here is $k=0$. Furthermore we note that for $\rho_-(x)$
the constraint in (\ref{eq:PF}) is automatically satisfied. We
have not been able yet to solve the above eigenvalue problem
analytically, and have instead simply resorted to numerical
evaluation of the largest eigenvalue in (\ref{eq:PtoF}). This
shows that the largest eigenvalue is indeed found for $k=0$ and
$\rho_-(x)$.

Our analysis of the P$\to$SG$_b$ transition can be simplified if
we use the fact that (\ref{eq:PSGb}) is written strictly in terms
of inner products of the various unit-length vectors. This allows
us to choose a convenient basis, e.g. one where
$\bsigma^2=(0,0,1)$. Upon again using polar coordinates to do the
integrations over the sphere $S_2$, we find that for $d=3$ the
second order P$\to$SG transition is to be solved from
\begin{eqnarray} \hspace*{-15mm}
\int_0^\pi\!\frac{\rmd\theta\sin(\theta)}{2}\int_{-\pi}^\pi\!\frac{\rmd\phi}{2\pi}
\int_0^\pi\!\frac{\rmd\theta^\prime\sin(\theta^\prime)}{2}\int_{-\pi}^\pi\!\frac{\rmd\phi^\prime}{2\pi}~
e^{\beta
J[\cos(\theta)+x\cos(\theta^\prime)+\sin(\theta^\prime)\cos(\phi^\prime)\sqrt{1-x^2}]}
\nonumber
\\
\hspace*{20mm} \times~
\psi(\sin(\theta)\sin(\theta^\prime)\cos(\phi)+\cos(\theta)\cos(\theta^\prime))
 =\frac{\sinh^2(\beta J)}{c(\beta J)^2} \psi(x) \nonumber
  \end{eqnarray}
Via suitable transformations of variables, viz. $t=\cos(\theta)$
and $s=\cos(\theta^\prime)$, and insertion of $\int\!\rmd
y~\delta[y-\sin(\theta)\sin(\theta^\prime)\cos(\phi)-\cos(\theta)\cos(\theta^\prime)]$
this expression can be simplified to
 \begin{eqnarray}
\hspace*{-15mm} \int_{-1}^1\!\frac{\rmd y \rmd s \rmd t}{4\pi}~
I_0( \beta J \sqrt{1-s^2}\sqrt{1-x^2})~ e^{\beta J[sx+t]}~
\frac{\theta[(1-s^2)(1-t^2)-(y-st)^2]}{\sqrt{(1-s^2)(1-t^2)-(y-st)^2}}
~\psi(y)\nonumber
\\ \hspace*{50mm}=\frac{\sinh^2(\beta J)}{c(\beta J)^2} \psi(x)
\label{eq:PSGind3}
  \end{eqnarray}
This latter equation is to be solved subject to the constraint
  $\int_{-1}^1\!\rmd y~\psi(y)=0$.
The integration kernel in (\ref{eq:PSGind3}) is again symmetric,
yielding as before either strictly symmetric or strictly
anti-symmetric eigenfunctions. Again the anti-symmetric
eigenfunctions offer the advantage of automatically satisfying the
appropriate constraint.\vsp

\begin{figure}[t]
\vspace*{5mm} \hspace*{5mm} \setlength{\unitlength}{0.75mm}
\begin{picture}(200,80)
\put(5,15){\includegraphics[height=60\unitlength,width=60\unitlength]
{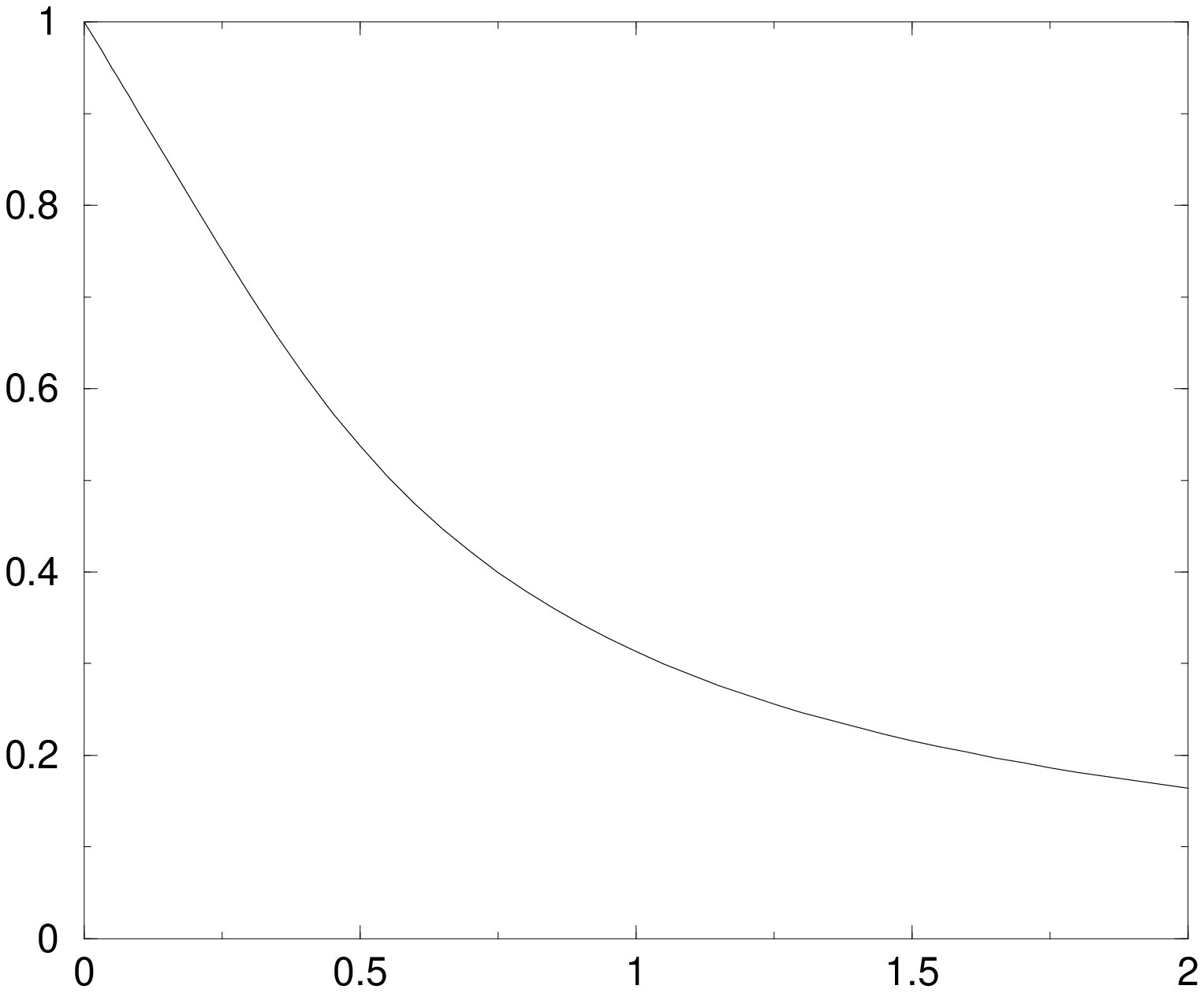}}
\put(72,15){\includegraphics[height=60\unitlength,width=60\unitlength]
{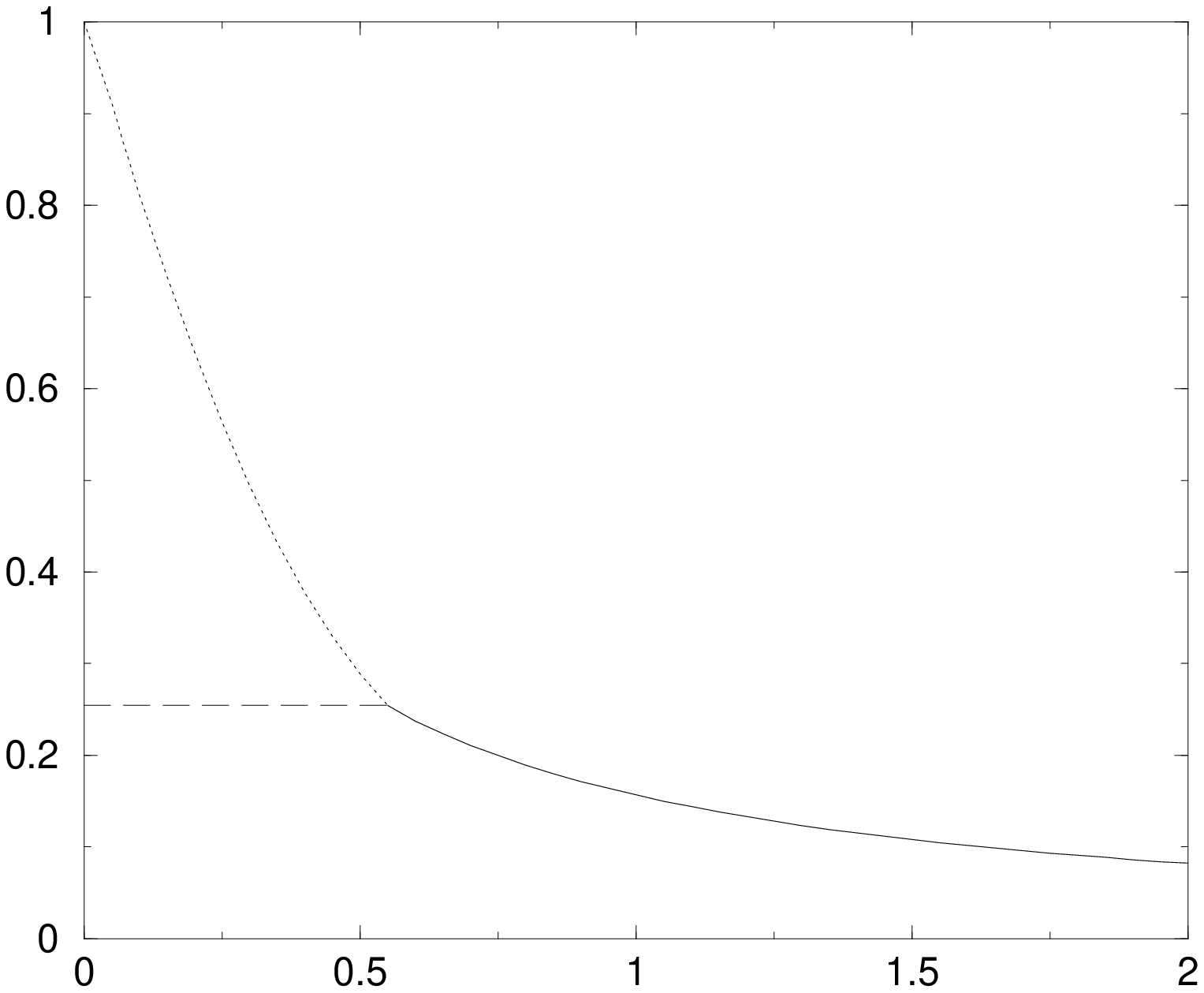}}
\put(139,15){\includegraphics[height=60\unitlength,width=60\unitlength]
{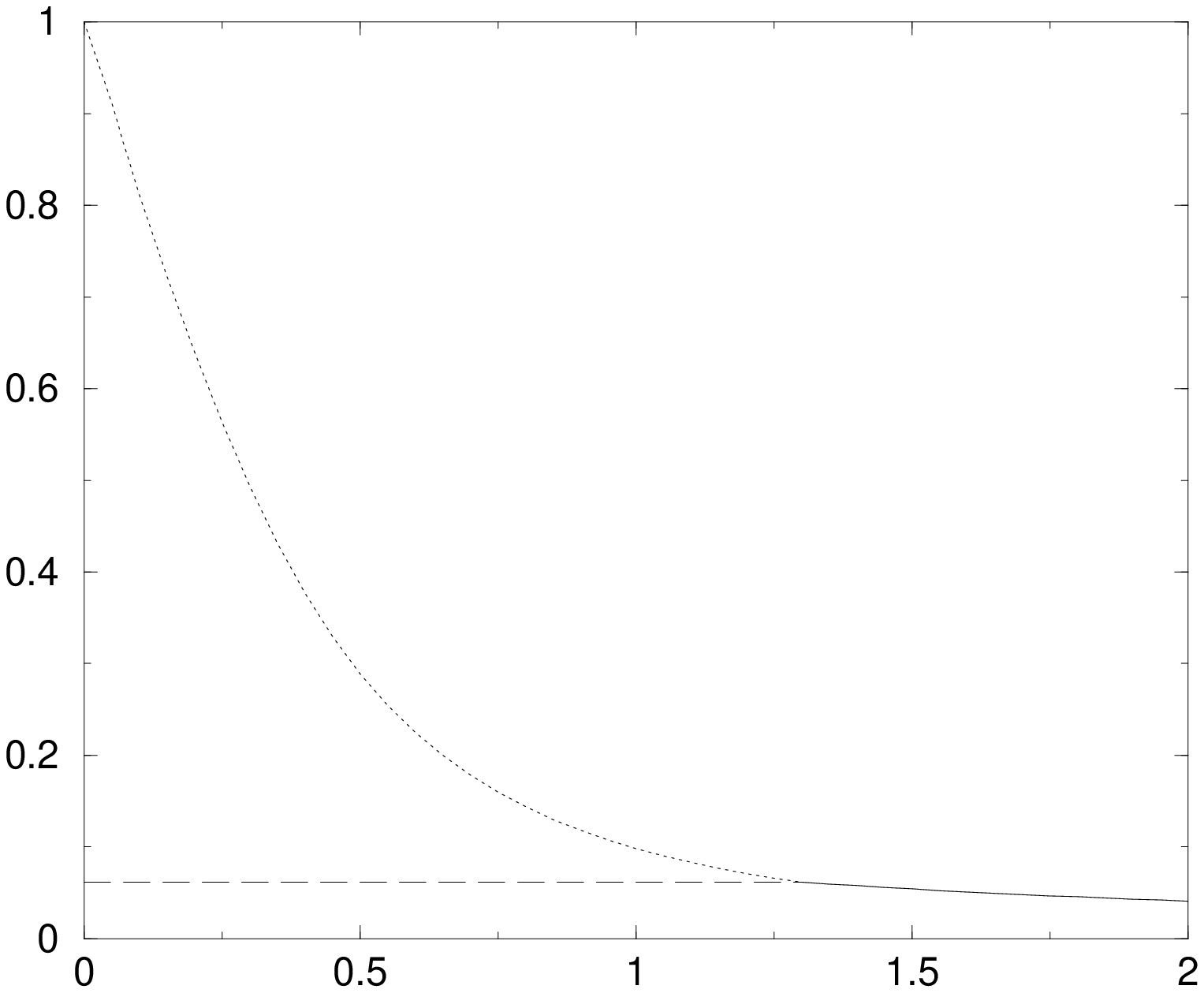}} \put(38,6){\here{\small
$T/J$}}\put(105,6){\here{\small $T/J$}}\put(172,6){\here{\small
$T/J$}} \put(0,45){\here{$c^{-1}$}} \put(41,80){\here{$\epsilon=1
$}}\put(108,80){\here{$\epsilon=\frac{1}{2}$}}
\put(175,80){\here{$\epsilon=\frac{1}{4}$}}
\put(48,60){\here{P}}\put(113,60){\here{P}} \put(182,60){\here{P}}
\put(21,30){\here{F}}
 \put(82,40){\here{SG}}
 \put(95,23){\here{F}}
 \put(149,35){\here{SG}}\put(170,20){\here{F}}
\end{picture}
\vspace*{-8mm} \caption{Phase diagrams in $d=3$, for the orthogonal
random matrix ensemble characterized by equation
(\ref{eq:ensemble_d3}), and for different values of $\epsilon$.
$\epsilon=1$ corresponds to purely ferromagnetic interactions,
whereas for $\epsilon=0$ they are fully random. As in the case
$d=2$, except for the location of the triple point, the SG$\to$F
transitions cannot be obtained from our present functional moment
expansions, but have been inferred from the assumption that there
is no change of phase character after the onset of order, when the
temperature is decreased further for fixed connectivity $c$.}
\label{figd3}
\end{figure}

We are now in a position to construct phase diagrams for $d=3$ and the
orthogonal random matrix ensemble characterized by
(\ref{eq:ensemble_d3}), by solving the remaining two eigenvalue
problems (\ref{eq:PtoFagain},\ref{eq:PSGind3}) with their associated
constraints numerically. We have done this for three different values
of $\epsilon$, leading to the phase diagrams in figure \ref{figd3},
which can be compared to the $d=2$ results of figure
\ref{fig:diagrams_binary}. As might be expected on physical grounds,
the extra degrees of freedom available to each spin in $d=3$ (which
increase the potential for the system to minimize its free energy
entropically, as opposed to energetically) lead to a lower transition
temperature to an ordered state, be it spin glass or ferromagnetic.
\begin{figure}[t]
\vspace*{15mm} \hspace*{5mm} \setlength{\unitlength}{0.71mm}
\begin{picture}(200,80)
 \put(70,9){\here{$T$}} \put(15,57){\here{$q,m$}}\put(167,9){\here{$c^{-1}$}}
\put(25,15){\includegraphics[height=83\unitlength,width=88\unitlength]
{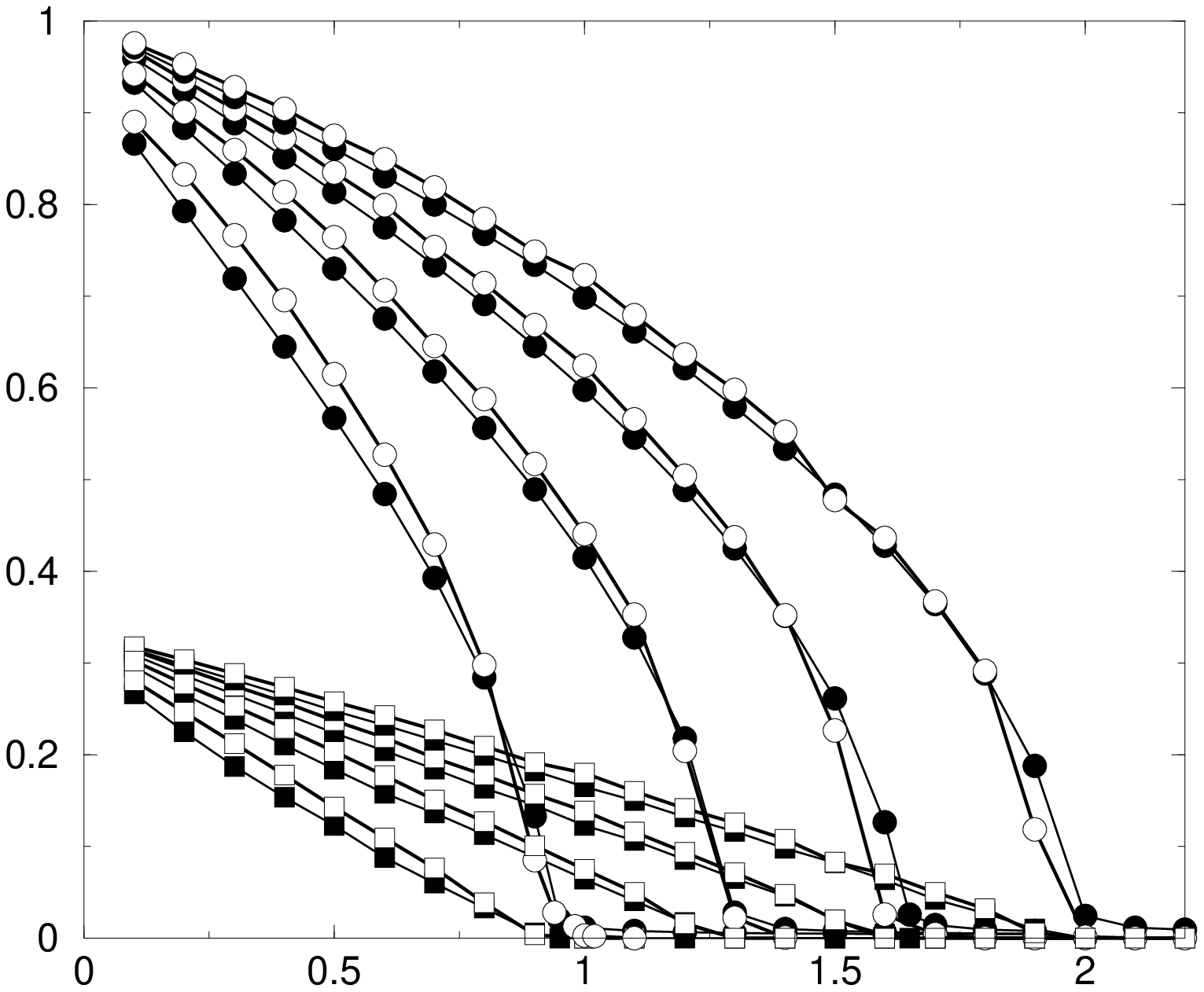}}
\put(119,15){\includegraphics[height=83\unitlength,width=88\unitlength]
{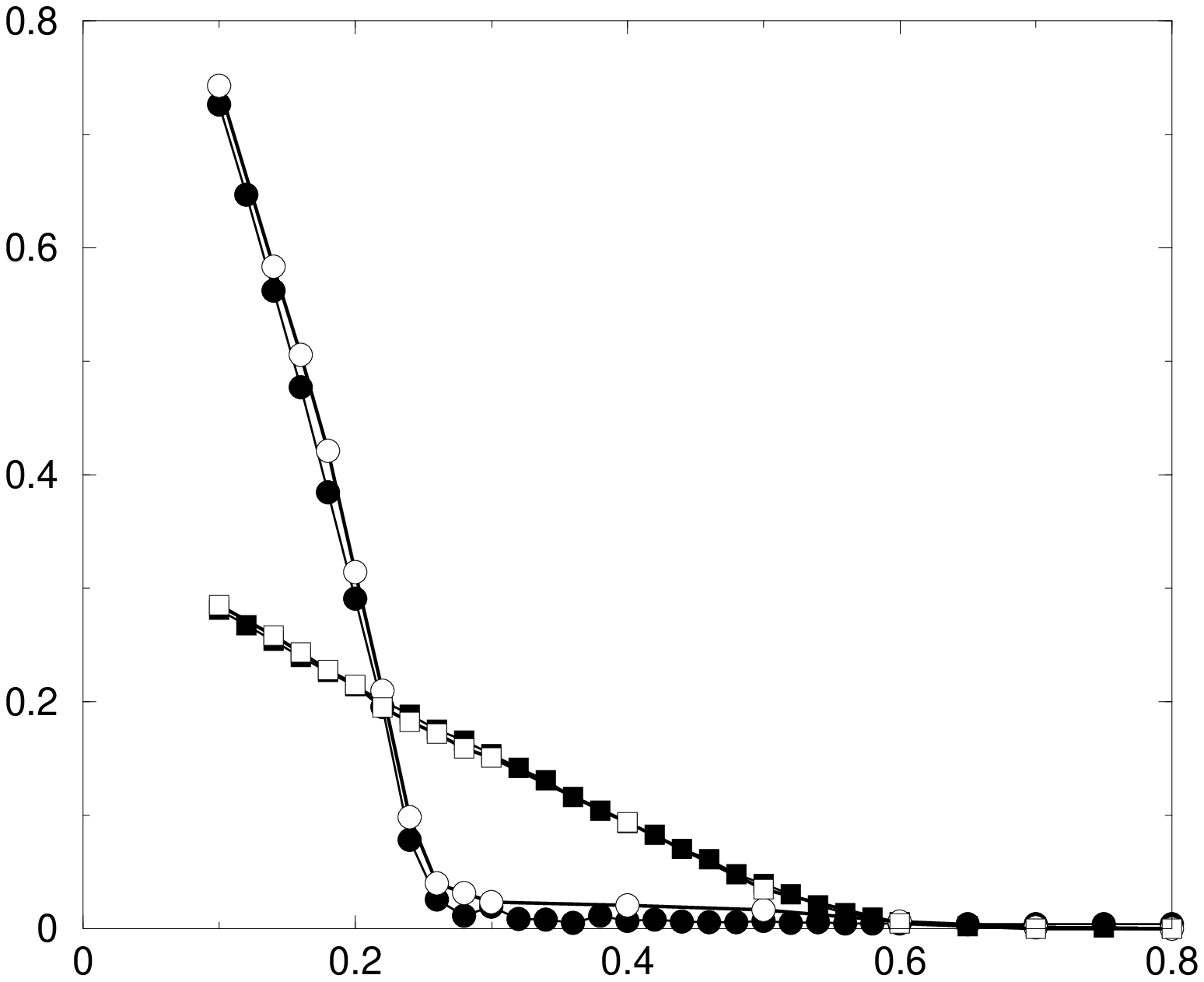}}
\end{picture}
\vspace*{-5mm} \caption{Comparison between theoretical
predictions, population dynamics (open symbols) and Monte Carlo
simulations (solid symbols) for $d=3$. Left picture: we show the
magnetization $m$ (circles) and the spin-glass order parameter $q$
(squares) defined in (\ref{eq:mq_d3}), for $\epsilon=1$ and for
connectivity values $c=3,4,5,6$ (from left to right). The
agreement indicates that our truncation in the parametrization of
(\ref{eq:d3_representation}) does not have a significant impact on
the numerical accuracy. Right picture: the order parameters $m$
(circles) and $q$ (squares) for $\epsilon=1/2$, along the line
$T=0.25$. The magnetization is seen to become zero around
$c^{-1}=0.25$,  thus verifying our assumption for the location of
the SG-F line.  All simulations were done for $N=10^5$
spins.}\label{fig:simul_d3}
\end{figure}

%%%%%%%%%%%%%%%%%%%%%%%%%%%%%%%%%%%%%%%%%%%%%%%%%%%%%%%%%%%%%%%%%%%%%%%%%%%%%%%%%%%%%%%%
%%%%%%%%%%%%%%%%%%%%%%%%%%%%%%%%%%%%%%%%%%%%%%%%%%%%%%%%%%%%%%%%%%%%%%%%%%%%%%%%%%%%%%%%
\subsection{Numerical calculation of order parameters via population dynamics}
%%%%%%%%%%%%%%%%%%%%%%%%%%%%%%%%%%%%%%%%%%%%%%%%%%%%%%%%%%%%%%%%%%%%%%%%%%%%%%%%%%%%%%%%
%%%%%%%%%%%%%%%%%%%%%%%%%%%%%%%%%%%%%%%%%%%%%%%%%%%%%%%%%%%%%%%%%%%%%%%%%%%%%%%%%%%%%%%%

Let us now turn to the numerical evaluation of the order
parameters in our system. In a spirit similar to that of the case
$d=2$, i.e. section \ref{sec:PDd2}, we will extract the relevant
population dynamics equations from
(\ref{eq:population_dynamics_specific}) upon making a suitable
choice for the parametrization of $P[\bsigma|\bmu]$. Here we will
again use the family of Fourier modes
\begin{eqnarray}
\fl \lefteqn{P(\phi,\theta|\bmu)= \frac{1}{D(\bmu)}\
e^{\sum_{m\geq 1}\left[A^c_m\cos(m\phi)+A^s_m\sin(m\phi)
+B^c_{m}\cos(m\theta)+B^s_{m}\sin(m\theta)\right]}}
\label{eq:d3_representation}
\\
& & \hspace*{-23mm} \times e^{\sum_{m,m^\prime\geq
1}\left[H^{cc}_{mm^\prime}\,\cos(m\phi)\cos(m^\prime\theta)+H_{mm^\prime}^{cs}\,\cos(m\phi)\sin(m^\prime\theta)+
H_{mm^\prime}^{sc}\,\sin(m\phi)\cos(m^\prime\theta)+H_{mm^\prime}^{ss}\,\sin(m\phi)\sin(m^\prime\theta)\!\right]}
\nonumber
\end{eqnarray}
where $D(\bmu)$ is the relevant normalization constant and $\bmu$
denotes collectively all coefficients
$\{A^\star,B^\star,H^{\star\star}\}$ with $\star\in\{c,s\}$. From
this point, one can proceed further by working out expression
(\ref{eq:population_dynamics_specific}) and converting it into one
for the coefficients $\{A^\star,B^\star,H^{\star\star}\}$. We have
implemented this strategy on the basis of a truncation after 8
coefficients, i.e.\@ we have taken
$A_m^\star\!=\!B_m^\star\!=\!\delta_{m,1}$ and
$H_{mm^\prime}^{\star\star}\!=\!\delta_{m,1}\delta_{m^\prime,1}$
so that $\bmu=(A^c,A^s,B^c,B^s,H^{cc},H^{cs},H^{sc},H^{ss})$
(details of the resulting expressions can be found in
\ref{sec:app_PDd3}). The self-consistent equation for $w(\bmu)$
that follows can then be solved using a population dynamics
prescription.

Finally, given the stationary distribution of coefficients
$w(\bmu)$ in the population dynamics method and expression
(\ref{eq:d3_representation}) one can then evaluate the order
parameters of the system, e.g.\@ the magnetization and the
spin-glass order parameter
\begin{equation}
m=\sqrt{m_x^2+m_y^2+m_z^2}
\hspace{10mm}
q=\frac13(q_x+q_y+q_z)
\label{eq:mq_d3}
\end{equation}
where now
\begin{eqnarray}
m_x=\Bra\Bra \cos(\phi)\sin(\theta)\Ket_{\phi,\theta|\bmu}\Ket_{\bmu}
& &\hspace{10mm}  q_x=\Bra\Bra \cos(\phi)\sin(\theta)\Ket_{\phi,\theta|\bmu}^2\Ket_{\bmu}
\\
m_y=\Bra\Bra \sin(\phi)\sin(\theta)\Ket_{\phi,\theta|\bmu}\Ket_{\bmu}
& &\hspace{10mm}   q_y=\Bra\Bra \sin(\phi)\sin(\theta)\Ket_{\phi,\theta|\bmu}^2\Ket_{\bmu}
\\
m_z=\Bra\Bra \cos(\theta)\Ket_{\phi,\theta|\bmu}\Ket_{\bmu}
& &\hspace{10mm}   q_z=\Bra\Bra \cos(\theta)\Ket_{\phi,\theta|\bmu}^2\Ket_{\bmu}
\end{eqnarray}
and with the averages
\begin{eqnarray}
\Bra (\cdots)\Ket_{\phi,\theta|\bmu}=\int_{-\pi}^\pi\!
\rmd\phi\int_{0}^\pi\! \rmd\theta\ |\sin(\theta)|\
P(\phi,\theta|\bmu)\ (\cdots)
\\
\Bra (\cdots)\Ket_{\bmu}=\int\! \rmd\bmu\ w(\bmu)\ (\cdots)
\end{eqnarray}

In figure \ref{fig:simul_d3} we show the results of evaluating our
observables numerically via the above strategy.  We compare our
bifurcation analysis (leading to predictions for the locations of
phase transitions) with the outcome of population dynamics
analysis and with data measured in simulation experiments. In
practice, our population dynamics shows good convergence already
for relatively modest population sizes of \@ $2000$ fields.

The Monte Carlo simulations were done in the present  case $d=3$
using the heat-bath algorithm of \cite{Miyatake1986,Loison2004},
and with system sizes of $10^5$ spins. To generate an ensemble of
uniformly distributed random orthogonal matrices it is convenient
to represent the rotation matrices using Euler angles
$(\alpha,\beta,\gamma)$ (in standard notation), i.e.
$\bU=R_z(\alpha)\,R_y(\beta)\,R_z(\gamma)$ for any $\bU\in SO(3)$.
Then, uniform integration over the Lie group $SO(3)$ is given by
the Haar measure $\rmd g_H(\alpha,\beta,\gamma)$ which, in Euler
representation, takes the form $\rmd
g_H(\alpha,\beta,\gamma)=(8\pi^2)^{-1}
\rmd\alpha\rmd\beta\rmd\gamma \sin(\beta)$, with
$\alpha,\gamma\in[0,2\pi]$ and $\beta\in[0,\pi]$. The relevant
matrix distribution $P(\bU)$ can then be written as
\begin{equation}
\fl
P(\bU)=\epsilon\delta[\bU-\one]+\frac{1-\epsilon}{8\pi^2}\int_0^{2\pi}\!\rmd\alpha\int_0^\pi\!
\rmd\beta\int_0^{2\pi}\!\rmd\gamma\ \sin(\beta)
\delta\left[\bU-R_z(\alpha)\,R_y(\beta)\,R_z(\gamma)\right]
\end{equation}
with
 \bd
 \fl
 {\footnotesize
R_z(\alpha)=\!\left(\!\begin{array}{ccc} \cos\alpha & \sin\alpha &
0
\\ -\sin\alpha & \cos\alpha & 0 \\ 0 & 0 & 1
\end{array}\!\right)~~~~
R_y(\beta)=\!\left(\!\begin{array}{ccc} \cos\beta & 0 & -\sin\beta
\\ 0 & 1 & 0   \\ \sin\beta & 0 & \cos\beta
\end{array}\!\right)
~~~~ R_z(\gamma)=\!\left(\!\begin{array}{ccc} \cos\gamma &
\sin\gamma & 0 \\ -\sin\gamma & \cos\gamma & 0 \\ 0 & 0 & 1
\end{array}\!\right)}
\ed \vsp

In the left picture of figure \ref{fig:simul_d3} we show the
magnetization and spin-glass order parameters (\ref{eq:mq_d3}) for
the case where $\epsilon=1$, i.e.\@  where our rotation matrices
reduce to the unit matrix $\bU=\one$ and the only source of
disorder in the system is the random nature of  the connectivity
variables $\{c_{ij}\}$. Here both the population dynamics and the
simulation experiments reproduce the P-F transition at the
predicted connectivity value. The excellent agreement obtained
indicates that, in retrospect, our truncation of the
parametrization (\ref{eq:d3_representation}) has been made with
only a minor cost in accuracy.  In the more involved scenarios
where $\epsilon<1$, a spin-glass phase with $m=0$ and $q>0$ will
become possible. An example is shown in the right picture of
figure \ref{fig:simul_d3}, where we chose the values
$\epsilon=1/2$ and $T=0.25$. For these parameter values our theory
predicts a P-SG transition, but also, based on physical grounds
(i.e. absence of re-entrance phenomena) we have assumed that the
elusive F-SG transition is located at the line segment parallel to
the $T$-axis, connecting the triple point where all phases meet to
the point $T/J=0$. In figure \ref{fig:simul_d3} we see that also
this latter assumption is verified numerically (by the population
dynamics results), and that the simulation data are, in turn,
again in good agreement with those of the population dynamics.

\section{Discussion}

In this paper we have applied the equilibrium replica method as
developed for finitely connected scalar spin systems to   models
of finitely connected unit-length vectorial spins of dimension
$d$, with pair-interactions which are given by random orthogonal
$d\times d$  matrices. Since our spins are continuous and the
connectivity $c$ is finite, rather than an effective field
distribution (as would have been the case for finitely connected
discrete scalar spins), here the replica-symmetric order parameter
is a functional. This generates a number of technical
complications. Firstly, rather than finding continuous transitions
away from the paramagnetic state by expansion of the RS order
parameter in powers of (scalar) moments of a field distribution,
here we have to generalize this procedure to an expansion in
functional moments. Secondly, one should expect serious numerical
difficulties when attempting to solve numerically the RS order
parameter functional from the appropriate self-consistent
nonlinear population dynamics equation. Here, however, these
difficulties  are kept under control due to the constrained nature
of the microscopic degrees of freedom. Since the present
continuous spins live on the sphere $S_{d-1}$, their value domain
is bounded, and it is therefore possible to construct efficient
and relatively accurate low-dimensional parametrizations of the
order parameter functional (in contrast to the case of unbounded
value domains, as e.g. with ordinary soft spins).

We have developed our theory initially for arbitrary values of the
dimension $d$ of the spins, and arbitrary choices of the ensemble
of random orthogonal matrices.  However, we ultimately calculate
phase diagrams and the values of moments of the order parameter
explicitly for $d=2$ (where our models reduce to finitely
connected XY spins with random chiral interactions) and for $d=3$
(where they reduce to finitely connected classical Heisenberg
spins with random chiral interactions). For $d=2,3$ we find three
types of phases: a paramagnetic (P), a ferromagnetic (F), and a
spin-glass phase (SG). The calculation of all continuous P$\to$F
and P$\to$SG transition lines can in all cases be reduced to the
(numerical) solution of relatively simple functional eigenvalue
problems; the F$\to$SG transition is constructed from the location
of the triple point, in combination with the conjecture (based on
previous experience with similar systems, and in line with the
Parisi-Toulouse hypothesis \cite{parisitoulouse} for the RSB
solution) that the phase entered at the onset of order (upon
leaving the paramagnetic state) will continue to hold upon
lowering the temperature further for fixed connectivity $c$. The
calculation of observables in the F and SG phases was carried out
using population dynamics techniques, applied to (truncated)
parametrizations of the order parameter functional, and the
results were tested against numerical simulations to reveal
excellent agreement.

We believe the main deliverables of this study to be the
successful extension and application of  finite connectivity
replica techniques to more demanding scenarios, where the
microscopic equilibrated degrees of freedom are neither discrete
nor of a scalar nature, and where also their interactions are of a
mathematically more complicated  form than just weighted (inner)
products. These techniques, which could easily be adapted to
non-Poissonnian random connectivity graphs,  are not only useful
tools in the study of physical systems, but may also be helpful to
analytically determine e.g. the influence of topology on global
processes in non-physical systems with scale-free connectivity,
such as the synchronization of randomly and finitely connected
planar oscillators \cite{jalan}.

\section *{Acknowledgment}

This study was initiated during an informal Finite Connectivity
Workshop at King's College London in November 2003. The authors
 acknowledge financial support from the
 the ESF SPHINX programme, the
Ministerio de Educaci\'{o}n, Cultura y Deporte (Spain, grant
SB2002-0107),  the Engineering and Physical Sciences Research
Council, the MCYT (Spain, grant BFM2003-08258), the FOM Foundation
(Fundamenteel Onderzoek der Materie, the Netherlands), and the
State Scholarships Foundation (Greece).

\appendix
\section{Details of population dynamics for $d=3$}
\label{sec:app_PDd3}

To arrive at a numerically tractable form of the population
dynamics relations, one is required to truncate
(\ref{eq:d3_representation}) to a relatively small number of
coefficients. Here we took
\begin{eqnarray}
\fl
\lefteqn{P(\phi,\theta|\bmu)= \frac{1}{D(\bmu)}\
\exp\left[A^c\cos(\phi)+A^s\sin(\phi)
+B^c\cos(\theta)+B^s\sin(\theta)\right]}
\\
& &
\fl
\times\
\exp\left[H^{cc}\,\cos(\phi)\cos(\theta)+H^{cs}\,\cos(\phi)\sin(\theta)+
H^{sc}\,\sin(\phi)\cos(\theta)+H^{ss}\,\sin(\phi)\sin(\theta)\right]
\nonumber
\end{eqnarray}
with $D(\bmu)$ the appropriate normalization constant. The update
relations for the above coefficients follow from the orthogonality
relations of the trigonometric functions:
\begin{eqnarray}
&& \fl w(\bmu)= \sum_{\ell\geq 0}\frac{e^{-c}c^{\ell}}{\ell!} \int
[\prod_{k\leq \ell} \rmd\bmu_k w(\bmu_k) \rmd\bU_k\, P(\bU_k)]
\label{eq:PD_d3}
\\
&& \hspace{-20mm} \times \delta\Big[A^c-\sum_{k\leq \ell}\frac{\pi
K_{c}(\bmu_k,\bU_k)-4 M_{cs}(\bmu_k,\bU_k)}{\pi(\pi^2-8)}\Big]\
\delta\Big[A^s-\sum_{k\leq \ell}\frac{\pi K_{s}(\bmu_k,\bU_k)-4
M_{ss}(\bmu_k,\bU_k)}{\pi(\pi^2-8)}\Big] \nonumber
\\
&& \hspace{-20mm} \times \delta\Big[B^c-\sum_{k\leq
\ell}\frac{1}{\pi^2}V_{c}(\bmu_k,\bU_k)\Big]\
\delta\Big[B^s-\sum_{k\leq \ell}\frac{\pi V_{s}(\bmu_k,\bU_k)-2
M(\bmu_k,\bU_k)}{\pi(\pi^2-8)}\Big] \nonumber
\\
&& \hspace{-20mm} \times \delta\Big[H^{cc}-\sum_{k\leq
\ell}\frac{2}{\pi^2}M_{cc}(\bmu_k,\bU_k)\Big]\
\delta\Big[H^{cs}-\sum_{k\leq \ell}2\frac{\pi
M_{cs}(\bmu_k,\bU_k)-2 K_{c}(\bmu_k,\bU_k)}{\pi(\pi^2-8)}\Big]
\nonumber
\\
&& \hspace{-20mm} \times \delta\Big[H^{sc}-\sum_{k\leq
\ell}\frac{2}{\pi^2}M_{sc}(\bmu_k,\bU_k)\Big]\
\delta\Big[H^{ss}-\sum_{k\leq \ell}2\frac{\pi
M_{ss}(\bmu_k,\bU_k)-2 K_{s}(\bmu_k,\bU_k)}{\pi(\pi^2-8)}\Big]
\nonumber
\end{eqnarray}
with the abbreviations
\begin{eqnarray}
 K_{c}(\bmu_k,\bU_k)&=&\int_{-\pi}^\pi\! \rmd\phi\ \cos(\phi)
\int_0^\pi\! \rmd\theta\ R(\phi,\theta;\bmu_k,\bU_k) \nonumber
\\
 K_{s}(\bmu_k,\bU_k)&=&\int_{-\pi}^\pi\! \rmd\phi\ \sin(\phi)
\int_0^\pi\! \rmd\theta\ R(\phi,\theta;\bmu_k,\bU_k) \nonumber
\\
 V_{c}(\bmu_k,\bU_k)&=&\int_{-\pi}^\pi\! \rmd\phi \int_0^\pi\!
\rmd\theta\ \cos(\theta)\ R(\phi,\theta;\bmu_k,\bU_k) \nonumber
\\
 V_{s}(\bmu_k,\bU_k)&=&\int_{-\pi}^\pi\! \rmd\phi \int_0^\pi\!
\rmd\theta\ \sin(\theta)\ R(\phi,\theta;\bmu_k,\bU_k)\nonumber
\nonumber
\\
 M_{cc}(\bmu_k,\bU_k)&=&\int_{-\pi}^\pi\! \rmd\phi\ \cos(\phi)
\int_0^\pi\! \rmd\theta\ \cos(\theta)\ R(\phi,\theta;\bmu_k)
\nonumber
\\
 M_{cs}(\bmu_k,\bU_k)&=&\int_{-\pi}^\pi\! \rmd\phi\ \cos(\phi)
\int_0^\pi\! \rmd\theta\ \sin(\theta)\ R(\phi,\theta;\bmu_k,\bU_k)
\nonumber
\\
 M_{sc}(\bmu_k,\bU_k)&=&\int_{-\pi}^\pi\! \rmd\phi\ \sin(\phi)
\int_0^\pi\! \rmd\theta\ \cos(\theta)\ R(\phi,\theta;\bmu_k,\bU_k)
\nonumber
\\
 M_{ss}(\bmu_k,\bU_k)&=&\int_{-\pi}^\pi\! \rmd\phi\ \sin(\phi)
\int_0^\pi\! \rmd\theta\ \sin(\theta)\ R(\phi,\theta;\bmu_k,\bU_k)
\nonumber
\\
 M(\bmu_k,\bU_k)&=&\int_{-\pi}^\pi\! \rmd\phi \int_0^\pi\!
\rmd\theta\ R(\phi,\theta;\bmu_\ell,\bU_k)\nonumber
\end{eqnarray}
\begin{eqnarray}
\hspace*{-10mm}
 R(\phi,\theta;\bmu,\bU)& = & \log(2\pi)+ \log\left[
\int_{0}^\pi\! \rmd\theta^\prime\ |\sin(\theta^\prime)|\
e^{B^c\cos(\theta^\prime)+B^s\sin(\theta^\prime)}\right.
\\
\hspace*{-10mm} & &  \times\ e^{\beta
J\left(U_{13}\cos(\theta^\prime)\sin(\theta)\cos(\phi)
+U_{23}\cos(\theta^\prime)\sin(\theta)\sin(\phi)+U_{33}\cos(\theta^\prime)\cos(\theta)\right)}
\nonumber
\\
\hspace*{-10mm}
 & & \times\ \left. I_0\left(
\sqrt{L_a^2(\phi,\theta,\theta^\prime,\bmu,\bU)+L_b^2(\phi,\theta,\theta^\prime,\bmu,\bU)}\right)
\right] \nonumber
\end{eqnarray}
Here $I_0(x)$ is the zero-th order modified Bessel function and
\begin{eqnarray}
\hspace*{-20mm}
 L_a(\phi,\theta,\theta^\prime,\bmu,\bU) & = &
A^c+H^{cc}\cos(\theta^\prime)+H^{cs}\sin(\theta^\prime) \nonumber
\hspace*{-20mm}
\\
\hspace*{-20mm}
 & &\hspace{-20mm}+ \beta J\Big[
U_{11}\sin(\theta)\cos(\phi)\sin(\theta^\prime)+U_{21}\sin(\theta)\sin(\phi)\sin(\theta^\prime)+U_{31}\cos(\theta)
\sin(\theta^\prime)\Big]\nonumber
\\
\hspace*{-20mm}
 L_b(\phi,\theta,\theta',\bmu,\bU) & = &
A^s+H^{sc}\cos(\theta')+H^{ss}\sin(\theta') \nonumber
\\
\hspace*{-20mm} & &\hspace{-20mm}+ \beta J\Big[
U_{12}\sin(\theta)\cos(\phi)\sin(\theta')+U_{22}\sin(\theta)\sin(\phi)\sin(\theta')+U_{32}\cos(\theta)\sin(\theta')\Big]\nonumber
\end{eqnarray}
\end{document}